\documentclass[11pt]{amsart}

\usepackage{graphicx,amssymb} 
\usepackage{epsfig}
\usepackage{mathrsfs}

\title[Fractal zeta functions and complex dimensions]{Fractal zeta functions and complex dimensions of relative fractal drums}

\dedicatory{\begin{center}To Professor Ha\" im Brezis, with profound admiration,\\
on the occasion of his 70th birthday.\\
Avec mes sinc\` eres remerciements, pour votre amiti\' e, vos enseignements,\\ et votre exemple \' edifiant. Votre \' etudiant, Michel.\end{center}}

\author{Michel\ L.\ Lapidus$^*$}
\thanks{$^*$The work of Michel L.\ Lapidus was partially supported by the US National Science Foundation (NSF) under the research grants DMS-0707524 and DMS-1107750, as well as by the Institut des Hautes Etudes Scientifiques (IHES) in Paris/Bures-sur-Yvette, France, where the first author was a visiting professor in the Spring of 2012 while part of this work was completed.}

\author{Goran\ Radunovi\'c}

\author{Darko\ \v Zubrini\'c$^{\dag}$}

\thanks{$^{\dag}$Goran Radunovi\'c and Darko \v{Z}ubrini\'c express their gratitude to the Ministry of Science of the Republic of Croatia for its support, as well as to the University of Zagreb for supporting a visit of the second author to the University of California, Riverside, in the Spring of 2014 during which a part of this article was completed.}

\begin{document}

\begin{abstract}
The theory of `zeta functions of fractal strings' has been initiated
by the first author in the early 1990s, and developed jointly with his collaborators during almost two decades of intensive research in numerous articles and several monographs.
In 2009, the same author introduced a new class of zeta functions, called `distance zeta functions', which since then, has enabled us to extend the existing theory of zeta functions of fractal strings and sprays to arbitrary bounded (fractal) sets in Euclidean spaces of any dimension.
A natural and closely related tool for the study of distance zeta functions is the class of `tube zeta functions', defined using the tube function 
of a fractal set. These three classes of zeta functions, under the name of `fractal zeta functions', exhibit deep connections with Minkowski contents 
and upper box dimensions, as well as, more generally, with the complex dimensions of fractal sets. Further extensions include zeta functions of relative fractal drums, the box dimension of which can assume negative values,
including minus infinity.
We also survey some results concerning the existence of the meromorphic extensions of the spectral zeta functions of fractal drums, based in an essential way on earlier results of the first author on the spectral (or eigenvalue) asymptotics of fractal drums.
It follows from these results that the associated spectral zeta function has a (nontrivial) meromorphic extension, and we use some of our results about fractal zeta functions to show the new fact according to which the upper bound obtained for the corresponding abscissa of meromorphic convergence is optimal. 
Finally, we conclude this survey article by proposing several open problems and directions for future research in this area.
\end{abstract}


\date{}

\maketitle

\tableofcontents

\newtheorem{theorem}{Theorem}[section]
\newtheorem{cor}[theorem]{Corollary}
\newtheorem{prop}[theorem]{Proposition}
\newtheorem{lemma}[theorem]{Lemma}

\theoremstyle{remark}
\newtheorem{remark}[theorem]{Remark}
\newtheorem{defn}[theorem]{Definition}
\newtheorem{example}[theorem]{Example}
\newtheorem{problem}[theorem]{Problem}
\newtheorem{exercise}[theorem]{Exercise}

\newcommand{\re}{\operatorname{Re}}
\newcommand{\im}{\operatorname{Im}}

\font\csc=cmcsc10

\def\esssup{\mathop{\rm ess\,sup}}
\def\essinf{\mathop{\rm ess\,inf}}
\def\wo#1#2#3{W^{#1,#2}_0(#3)}
\def\w#1#2#3{W^{#1,#2}(#3)}
\def\wloc#1#2#3{W_{\scriptstyle loc}^{#1,#2}(#3)}
\def\osc{\mathop{\rm osc}}
\def\var{\mathop{\rm Var}}
\def\supp{\mathop{\rm supp}}
\def\Cap{{\rm Cap}}
\def\norma#1#2{\|#1\|_{#2}}

\def\C{\Gamma}

\let\text=\mbox

\catcode`\@=11
\let\ced=\c
\def\a{\alpha}
\def\b{\beta}
\def\d{\delta}
\def\g{\lambda}
\def\o{\omega}
\def\q{\quad}
\def\n{\nabla}
\def\s{\sigma}
\def\div{\mathop{\rm div}}
\def\sing{{\rm Sing}\,}
\def\singg{{\rm Sing}_\ty\,}

\def\A{{\cal A}}
\def\F{{\cal F}}
\def\H{{\cal H}}
\def\W{{\bf W}}
\def\M{{\cal M}}
\def\N{{\cal N}}
\def\S{{\cal S}}

\def\eR{{\bf R}}
\def\eN{{\bf N}}
\def\Ze{{\bf Z}}
\def\Qe{{\bf Q}}
\def\Ce{{\bf C}}

\def\ty{\infty}
\def\e{\varepsilon}
\def\f{\varphi}
\def\:{{\penalty10000\hbox{\kern1mm\rm:\kern1mm}\penalty10000}}
\def\ov#1{\overline{#1}}
\def\DD{\Delta}
\def\O{\Omega}
\def\pa{\partial}

\def\st{\subset}
\def\stq{\subseteq}
\def\pd#1#2{\frac{\pa#1}{\pa#2}}
\def\sgn{{\rm sgn}\,}
\def\sp#1#2{\langle#1,#2\rangle}

\newcount\br@j
\br@j=0
\def\q{\quad}
\def\gg #1#2{\hat G_{#1}#2(x)}
\def\inty{\int_0^{\ty}}
\def\od#1#2{\frac{d#1}{d#2}}

\def\bg{\begin}
\def\eq{equation}
\def\bgeq{\bg{\eq}}
\def\endeq{\end{\eq}}
\def\bgeqnn{\bg{eqnarray*}}
\def\endeqnn{\end{eqnarray*}}
\def\bgeqn{\bg{eqnarray}}
\def\endeqn{\end{eqnarray}}

\def\bgeqq#1#2{\bgeqn\label{#1} #2\left\{\begin{array}{ll}}
\def\endeqq{\end{array}\right.\endeqn}

\def\abstract{\bgroup\leftskip=2\parindent\rightskip=2\parindent
        \noindent{\bf Abstract.\enspace}}
\def\endabstract{\par\egroup}

\def\udesno#1{\unskip\nobreak\hfil\penalty50\hskip1em\hbox{}
             \nobreak\hfil{#1\unskip\ignorespaces}
                 \parfillskip=\z@ \finalhyphendemerits=\z@\par
                 \parfillskip=0pt plus 1fil}
\catcode`\@=11

\def\cal{\mathcal}
\def\eR{\mathbb{R}}
\def\eN{\mathbb{N}}
\def\Ze{\mathbb{Z}}
\def\Qu{\mathbb{Q}}
\def\Ce{\mathbb{C}}

\def\osd{\mathrm{osd}\,}

\def\sdim{\mbox{\rm s-dim}\,}
\def\sd{\mbox{\rm sd}\,}

\def\res{\operatorname{res}}

\def\L{{\cal L}}
\def\po{\mathcal{P}}
\def\I{\mathrm{i}}
\def\E{\mathrm{e}}
\def\D{\mathrm{d}}
\def\qs{\q}
\def\avM{\widetilde{\mathcal{M}}}

\newenvironment{xtolerant}[2]{%
  \par
  \ifx\empty#1\empty\else\tolerance=#1\relax\fi
  \ifx\empty#2\empty\else\emergencystretch=#2\relax\fi
}{%
  \par
}

\newpage

\section{Introduction}\label{intro}

The aim of this article is to provide a survey of some recent advances in the theory of fractal zeta functions, based on the monograph \cite{fzf} by the authors of the present work.

\medskip 

In 2009, it became clear that the theory of zeta functions of fractal strings, developed by the first author (jointly with several collaborators) beginning
in the early 1990s and described in an extensive joint monograph \cite{lapidusfrank12} with van Frankenhuijsen (see also the references therein), can be viewed from a much wider perspective.
This was due to the discovery (by the first author) of the new class of zeta functions, called `distance zeta functions',
associated with arbitrary bounded (fractal) sets in Euclidean spaces of any dimension; see Definition \ref{defn}.
This discovery opens new perspectives in the study of the complex dimensions of fractal sets.
As a result, in~[LapRa\v Zu1--7], 
the foundations for the theory of fractal zeta functions have now been laid.
Along with the corresponding (and closely related) tube zeta functions, introduced in Definition \ref{zeta_tilde} and developed as a valuable technical tool, it exhibits very interesting connections with the Minkowski contents and dimensions of fractal sets; see Theo\-rem~\ref{pole1mink_tilde}.

\medskip

The notion of a distance zeta function, as well as that of a tube zeta function, can be further extended to a new class of objects, called `relative fractal drums', which includes
fractal strings and bounded sets as well as standard fractal drums as special cases; see Section~\ref{relative}. An unexpected novelty is that the associated box dimensions of relative fractal drums may assume negative values (see Proposition \ref{ndim}), and even the value minus infinity (see Corollary \ref{flat}).
This new phenomenon reflects the intuitive picture of `flatness' of the relative fractal drum under consideration.

\medskip

The basic property of the distance zeta function of a fractal set is that its abscissa of (absolute or Lebesgue) convergence is equal to the upper box dimension of the set; see Theorem \ref{an}. Furthermore, assuming that this value
is a pole,\footnote{Under mild hypotheses, it is always a singularity; see part~$(c)$ of Theorem~\ref{an}.} then it is simple.
Moreover, the residue of the distance zeta function computed at the upper box dimension is closely related to the corresponding upper and lower Minkowski contents.
A similar statement holds for the tube zeta function.
(See Theorems~\ref{pole1} and~\ref{pole1mink_tilde}.)
In this way, these new fractal zeta functions (i.e., the distance and tube zeta functions) help build a bridge connecting the geometry of fractal sets with analytic number theory and complex analysis.

\medskip

We note that (as is stated in part $(b)$ of Theorem~\ref{an} and in Corollary~\ref{an1}), under mild additional assumptions, the abscissa of absolute convergence of the distance zeta function coincides not only with the upper box (or Minkowski) dimension $D$ of the given bounded set (or, more generally, of the relative fractal drum, see part $(b)$ of Theorem~\ref{an_rel} along with Corollary~\ref{an1r}), but also with the abscissa of holomorphic convergence of the zeta function: the open half-plane $\{\re s>D\}$ is both the maximal right open half-plane on which the Lebesgue integral initially defining the distance zeta function is convergent (and hence, absolutely convergent) and to which the zeta function can be holomorphically extended.
(See part $(c)$ of Theorem~\ref{an} along with Corollary~\ref{2.31/2}, in the case of bounded subsets of $\eR^N$, and their counterpart for general relative fractal drums, part $(c)$ of Theorem~\ref{an_rel} along with Corollary~\ref{2.31/2r}.)

\medskip

We stress that provided $D$ (the upper box dimension) is not equal to $N$ (the dimension of the ambient Euclidean space $\eR^N$), that is, if $D<N$ (since we always have $D\leq N$), all the results concerning the tube zeta functions have exact counterparts for the distance zeta functions (either for arbitrary bounded subsets of $\eR^N$ or, more generally, for relative fractal drums in $\eR^N$); see Remark~\ref{2.71/2} at the end of Section~\ref{disttube} below.
In particular, the tube and distance zeta functions can either be simultaneously meromorphically extended to a given domain $U$ of $\Ce$ (containing $\{\re s\geq D\}$) or else cannot be extended at all; furthermore, in the former case and if $D<N$, their meromorphic extensions then have the same poles in $U$ (i.e., they have the same {\em visible complex dimensions}) and their respective residues at $D$ are very simply related.\footnote{In the case when $D=N$, the statement about the poles in $U$ is also true except for the pole at $s=N$ which is then a pole of the tube zeta function and a removable singularity of the distance zeta function.}
For example, if the given bounded set $A$ is Minkowski measurable, then the residue of the tube zeta function at $D$ coincides with the Minkowski content of $A$; the exact analogue of this statement also holds for relative fractal drums. 
(See Remark~\ref{2.71/2} below along with~\cite{fzf} for detailed information.)
Accordingly, the fractal zeta functions introduced in [LapRa\v Zu1--7] 
contain essentially the same information.
However, for computational or theoretical purposes, and depending on the given situation or example under consideration, it is more convenient to use either the distance zeta function or the tube zeta function.
Many illustrations of this latter statement are provided throughout~\cite{fzf}.

\medskip

In Section~\ref{qp_sets}, we introduce the so-called transcendentally $n$-quasiperiodic sets for any integer $n\ge2$ (that is, roughly speaking, the sets possessing $n$ quasiperiods; see Definition \ref{quasiperiodic}), and describe their construction; see Theorems \ref{trans} and \ref{quasi1}. It is also possible to construct $\ty$-quasiperiodic sets, that is, sets which possess infinitely many quasiperiods; see Theorem \ref{ty}. The constructions are based on carefully chosen sequences of generalized Cantor sets with two parameters, introduced in Definition \ref{Cma}.

\medskip

In Section~\ref{relative}, as was mentioned above, we introduce the notion of a relative fractal drum, which significantly extends the classic notion of a fractal drum.
It provides a lot of useful additional flexibility to the (higher-dimensional) theory of complex dimensions, and gives rise to new and interesting phenomena (such as the lowering of the Minkowski dimension, to the point that the latter relative dimension can become negative and even equal $-\ty$, as in Proposition~\ref{ndim} and Corollary~\ref{flat}, respectively, which is clearly impossible in the standard theory).

\medskip

In Section~\ref{szf}, we briefly recall some of the key results of the first author in~\cite{Lap1} (see also [Lap2--3]) 
concerning the asymptotics of the eigenvalues (or, equivalently, the frequencies) of fractal drums in $\eR^N$, either for Dirichlet Laplacians or (under suitable hypotheses) for the Neumann Laplacians, as well as for higher order self-adjoint elliptic operators (with possibly variable coefficients and mixed Dirichlet--Neumann boundary conditions).
We deduce from these results the existence (in an appropriate open right half-plane) of a (necessarily unique) meromorphic extension of the spectral zeta function of a given fractal drum.
This result was already obtained in [Lap2--3] 
but we give a slightly more precise statement of it in~\cite{fzf} and Section~\ref{szf} below, as well as provide an elementary proof of the passage from the main error estimates of~\cite{Lap1} to the existence of a meromorphic extension of the spectral zeta function.
We also use the results of Section~\ref{relative} in an essential way (specifically, Theorem~\ref{ty}) in order to show that for the Dirichlet (or also the Neumann) Laplacian, there exist bounded open sets (or relative fractal drums) in $\eR^N$ such that the maximal half-plane of meromorphic convergence of the spectral zeta function actually coincides with the half-plane of (absolute) convergence of the associated fractal zeta functions, thereby proving the sharpness (or optimality) of the inequality
$$
D_{mer}(\zeta_{\O}^*)\leq D,
$$
where $D$ denotes the inner Minkowski dimension of $\partial\O$ (relative to $\O$).
This result is new even in dimension one (i.e., when $N=1$).

\medskip

In Section~\ref{class}, we provide a classification of bounded subsets of $\eR^N$, based on the results described in the rest of the paper.
In particular, we briefly discuss Minkowski degenerate, nondegenerate, and measurable (or constant) bounded sets, respectively.
We also discuss periodic and (algebraically or transcendentally) quasiperiodic subsets of $\eR^N$.
A more detailed discussion of these various categories of bounded sets can be found in [LapRa\v Zu1--7].

\medskip

Finally, in Section~\ref{open}, we provide several open problems motivated by the work in [LapRa\v Zu1--7] 
(and [Lap--vFr1--3]), 
and connecting various aspects of fractal geometry, geometric analysis, mathematical physics, harmonic analysis, dynamical systems, partial differential equations, spectral geometry and analytic number theory.
Many additional open problems and possible research directions are proposed in~\cite[Chapter~5]{fzf}.
However, the problems stated in the first half of Section~\ref{open} are original to the present survey article.
We hope that some readers will be inclined to address them.

\medskip

Let us now close this introduction by recalling some basic notation which will be used throughout this paper.
Assume that $A$ is a given bounded subset of $\eR^N$ and let $r$ be a fixed real number. We define the {\em upper} and {\em lower $r$-dimensional Minkowski contents of $A$}, respectively, by
$$
\M^{*r}(A):=\limsup_{t\to0^+}\frac{|A_t|}{t^{N-r}},\q \M_*^r(A):=\liminf_{t\to0^+}\frac{|A_t|}{t^{N-r}},
$$
where $A_t$ denotes the Euclidean $t$-neighborhood of $A$.
The {\em upper} and {\em lower box $(\mathrm{or}$ Minkowski$)$ dimensions of $A$} are defined, respectively, by
$$
\ov\dim_BA=\inf\{r\in\eR:\M^{*r}(A)=0\},\q \underline\dim_BA=\inf\{r\in\eR:\M_*^r(A)=0\}.
$$
It is always the case that $0\leq\underline{\dim}_BA\leq\ov{\dim}_BA\leq N.$
Furthermore, if $A$ is such that $\ov\dim_BA=\underline\dim_BA$, then this common value is denoted by $\dim_BA$ and is called the {\em box $(\mathrm{or}$ Minkowski$)$ dimension of $A$}.
Moreover, if $A$ is such that, for some $D\in[0,N]$, we have $0<\M_*^D(A)\le\M^{*D}(A)<\ty$ (in particular, then $\dim_BA$ exists and $D=\dim_BA$), we say that {\em $A$ is  Minkowski nondegenerate}. If $\M_*^D(A)=\M^{*D}(A)$, then the common value is denoted by $\M^D(A)$ and called the {\em Minkowski content of} $A$. Finally, assuming that $A$ is such that $\M^D(A)$ exists and $0<\M^D(A)<\ty$, we say that {\em $A$ is Minkowski measurable}.

\medskip

In closing, we note that the notion of Minkowski dimension was introduced (for noninteger values) by Bouligand in~\cite{Bou} in the late 1920s (without making a clear distinction between the lower and upper limits), while the notion of Minkowski content and measurability was introduced in~\cite{federer} and~\cite{stacho}, respectively.

\section{Distance and tube zeta functions}\label{disttube}

We introduce a new class of zeta functions, defined by the first author in 2009, which
extends the notion of geometric zeta functions of bounded fractal strings to bounded subsets of Euclidean spaces of arbitrary dimension.

\medskip

\begin{defn}[\cite{fzf}]\label{defn}  Let $\delta$ be a fixed positive real number.
Then, the {\em distance zeta function} $\zeta_A$ of a given bounded subset $A$ of $\eR^N$ is defined by
\begin{equation}\label{z}
\zeta_A(s):=\int_{A_\delta}d(x,A)^{s-N}\D x,
\end{equation}
for all $s\in\Ce$ with $\re s$ sufficiently large. Here, $d(x,A)$ denotes the Euclidean distance from $x$ to $A$.
Furthermore, the integral is taken in the sense of Lebesgue (and hence, is necessarily absolutely convergent).
\end{defn}

\medskip

\begin{remark}
The dependence of the distance zeta function $\zeta_A$ on $\delta>0$ is inessential, in the sense that we are mostly interested in the singularities of $\zeta_A$ (more precisely, in the singularities of meromorphic extensions of $\zeta_A$), which are preserved when varying $\delta$.
This follows from the fact that the difference of two distance zeta functions of the same set $A$ corresponding to two different values of $\delta$ is an entire function.
The same comment applies verbatim to the tube zeta function $\widetilde{\zeta}_A$ to be introduced in Definition \ref{zeta_tilde} below.  
\end{remark} 

\medskip

The distance zeta function represents a natural extension of the notion of a {\em geometric zeta function} $\zeta_{\mathcal{L}}$ of a bounded fractal string $\mathcal{L}=(l_j)_{j\ge1}$ (also introduced by the first author in the early 1990s and extensively studied in [Lap--vFr1--3] 
and in the relevant references therein):
\begin{equation}
\zeta_{\mathcal{L}}(s):=\sum_{j=1}^\ty l_j^s,
\end{equation}
for every $s\in\Ce$ with $\re s$ sufficiently large.
Here, by a {\em bounded fractal string} $\mathcal{L}$ we mean any nonincreasing sequence of positive real numbers $(l_j)_{j\ge1}$ such that $l:=\sum_{j\ge1}l_j<\ty$.
It is easy to see that if we define the set
$$
A_{\mathcal{L}}:=\bigg\{a_k:=\sum_{k\ge j}l_j:k\ge1\bigg\},
$$
then the distance zeta function
$\zeta_{A_{\mathcal{L}}}$ and the geometric zeta function $\zeta_{\mathcal{L}}$ are connected with the following relation:
\begin{equation}\label{dist_geo}
\zeta_{A_{\mathcal{L}}}(s)=f(s)\,\zeta_{\mathcal{L}}(s)+g(s),
\end{equation}
for all complex numbers $s$ such that $\re s$ is sufficiently large, where $f$ and $g$ are holomorphic on $\Ce\setminus\{0\}$ and $f$ is nowhere vanishing. 
In particular, due to Theorem \ref{an} below, it follows that the abscissae of convergence  of the distance zeta function $\zeta_A$ and 
of the geometric zeta function $\zeta_{\mathcal{L}}$ coincide, and that the corresponding sets of their poles on the critical line, as well as their multiplicities, also coincide. For more details, see \cite{fzf}.

\medskip

Throughout this paper, given $\alpha\in\eR\cup\{\pm\ty\}$, we will use the short-hand notation $\{\re s>\alpha\}$ to stand for the open right half-plane $\{s\in\Ce\,:\,\re s>\alpha\}$.
In particular, if $\alpha=+\ty$ or $\alpha=-\ty$, this half-plane becomes the empty set $\emptyset$ or the whole complex plane $\Ce$, respectively.
Similarly, if $\alpha\in\eR$, then the short-hand notation $\{\re s=\alpha\}$ stands for the vertical line $\{s\in\Ce\,:\,\re s=\alpha\}$.
If, in particular, $\alpha:=\ov{\dim}_BA$ $(=D(\zeta_A))$, the abscissa of convergence of $\zeta_A$ (see Corollary~\ref{an1} below and the text preceding it), where $A\subseteq\eR^N$ is bounded, then the vertical line is referred to as the {\em critical line} (of $\zeta_A$ or, simply, of $A$).
We will implicitly adopt an analogous terminology in the more general case of a relative fractal drum $(A,\O)$ discussed in Section~\ref{relative}.
Then, since (by Corollary~\ref{an1r}) $\ov{\dim}_B(A,\O)=D(\zeta_A(\,\cdot\,,\O))$, the vertical line $\{\re s=\ov{\dim}_B(A,\O)\}$ is referred to as the {\em critical line} (of $\zeta_A(\,\cdot\,,\O)$ or, simply, of $(A,\O)$).

\medskip

In the sequel, in order to avoid trivial special cases, we assume implicitly that all bounded subsets of $\eR^N$ under consideration in the statement of the theorems are nonempty. For example, in Theorem~\ref{an} and Corollaries~\ref{an1} and~\ref{2.31/2} below, the bounded set $A\subseteq\eR^N$ is implicitly assumed to be {\em nonempty}.

\medskip

The following result describes some of the basic properties of distance zeta functions.

\begin{theorem}[\cite{fzf}]\label{an} Let $A$ be an arbitrary bounded subset of $\eR^N$ and let $\delta>0$. Then$:$

\bigskip

$(a)$ The distance zeta function $\zeta_A$ defined by \eqref{z} is holomorphic in the 
half-plane $\{\re s>\ov\dim_BA\}$.

\bigskip

$(b)$ In part $(a)$, the half-plane $\{\re s>\ov\dim_BA\}$ is optimal, from the point of view of the convergence $($and therefore, also of the absolute convergence$)$ of the Lebesgue integral defining $\zeta_A(s)$ in~\eqref{z}$;$ i.e., the equality~\eqref{ddim} in Corollary~\ref{an1} below holds.
Furthermore, $\zeta_A(s)$ is still given by~\eqref{z} for $\re s>\ov{\dim}_BA;$ i.e., the inequality~\eqref{51/4} in Corollary~\ref{an1} below holds.


\bigskip

$(c)$ If the box $($or Minkowski$)$ dimension $D:=\dim_BA$ exists, $D<N$,\footnote{Since we always have $0\leq\ov{\dim}_BA\leq N$, assuming that $D=\dim_BA<N$ is equivalent to assuming that $D\neq N$.} and $\M_*^D(A)>0$, then $\zeta_A(s)\to+\ty$ as $s\in\eR$
converges to $D$ from the right.\footnote{Hence, $D$ is a singularity (which may or may not be a pole) of $\zeta_A$. Naturally, if $\zeta_A$ possesses a meromorphic continuation to an open neighborhood of $D$, then $D$ is a pole of $\zeta_A$. Section \ref{merom_ext} and Section \ref{relative} will provide several sufficient conditions under which $\zeta_A$ can be meromorphically continued beyond the critical line $\re s =D$,
and hence, in particular, to an open neighborhood of $D$.}
Therefore, the two equalities in~\eqref{51/2} of Corollary~\ref{2.31/2} below hold. 
\end{theorem}

\medskip

The proof of part $(a)$ of Theorem \ref{an} rests in part on the following interesting and little-known result, by Harvey\label{harvey} and Polking\label{polking}, stated implicitly on page 42 of \cite{acta}: 
\begin{equation}\label{int}
\mbox{If $\gamma\in(-\ty,N-\ov\dim_BA)$, then $\displaystyle\int_{A_\delta}d(x,A)^{-\gamma}\D x<\ty$,}\footnote{Assuming that $D=\dim_BA$ exists and $\mathcal{M}^D_*(A)>0$, then the converse implication holds as well; see \cite[Theorem 4.1]{proc05}.}
\end{equation}
where $\delta$ is any fixed positive real number. 
 Harvey and Polking used it in their study of removable singularities of solutions of certain partial differential equations.

\medskip

Before stating several consequences of Theorem~\ref{an} (namely, Corollaries~\ref{an1} and~\ref{2.31/2} below), we need to introduce some terminology and notation, which will also be used in the remainder of the paper.

\medskip

Given a meromorphic function (or, more generally, an arbitrary complex-valued function) $f=f(s)$, initially defined on some domain $U\subseteq\Ce$, we denote by $D_{hol}(f)$ the unique extended real number (i.e., $D_{hol}(f)\in\eR\cup\{\pm\ty\}$) such that $\{\re s>D_{hol}(f)\}$ is the {\em maximal} open right half-plane (of the form $\{\re s>\alpha\}$, for some $\alpha\in\eR\cup\{\pm\ty\}$) to which the function $f$ can be holomorphically extended.\footnote{By using the principle of analytic continuation, it is easy to check that $D_{hol}(f)$ is well defined; see~\cite[Section~2.1]{fzf}.}
This maximal (i.e., largest) half-plane is denoted by $\mathcal H(f)$ and called the {\em half-plane of holomorphic convergence of $f$}.

\medskip

If, in addition, the function $f=f(s)$ is assumed to be given by a {\em Dirichlet-type integral},\footnote{This is the case of the classic (generalized) Dirichlet series and integrals~\cite{serre,Pos}, as well as of the geometric zeta functions of fractal strings studied in [Lap--vFr1--3] 
and of all the fractal zeta functions considered in this paper and in [LapRa\v Zu1--5], 
including the distance and tube zeta functions ($\zeta_A$ and $\widetilde{\zeta}_A$) and their relative counterparts ($\zeta_A(\,\cdot\,,\O)$ and $\widetilde{\zeta}_A(\,\cdot\,,\O)$) for relative fractal drums discussed in Section~\ref{relative}; see~\cite[Section~2.1]{fzf}.
(It is also the case, in particular, of the spectral zeta functions of fractal drums discussed in Section~\ref{szf} below.)} of the form
\begin{equation}\label{41/4}
f(s)=\int_E\varphi(x)^s\D\mu(x),
\end{equation}
for $s\in\Ce$ with $\re s$ sufficiently large, where $\mu$ is a suitable (positive or complex) local (i.e., locally bounded) measure on a given (measurable) space $E$, $\varphi\geq 0$ $\mu$-a.e.\ on $E$ and $\varphi$ is $|\mu|$-essentially bounded from above, then $D(f)$, the {\em abscissa of $($absolute $\mathrm{or}$ Lebesgue$)$ convergence of} $f$, is the unique extended real number (i.e., $D(f)\in\eR\cup\{\pm\ty\}$) such that $\{\re s>D(f)\}$ is the {\em maximal} open right half-plane (of the form $\{\re s>\alpha\}$, for some $\alpha\in\eR\cup\{\pm\ty\}$) on which the Lebesgue integral initially defining $f$ in~\eqref{41/4} is convergent (or, equivalently, is absolutely convergent), with $\mu$ replaced by $|\mu|$, the total variation measure of $\mu$.
(Recall that $|\mu|=\mu$ if $\mu$ is positive.)
In short, $D(f)$ is called the {\em abscissa of convergence of} $f$.
Furthermore, the aforementioned maximal right half-plane is denoted by $\Pi(f)$ and is called the {\em half-plane of} (absolute or Lebesgue) {\em convergence of} (the Dirichlet-type integral) $f$.
It is shown in~\cite[Section~2.1 and Appendix~A]{fzf} that 
$D(f)$ is well defined and (with the notation of~\eqref{41/4} just above) we have, equivalently:\footnote{Let $D:=\ov{\dim}_BA$, for brevity.
In light of Theorem~\ref{an} and Corollary~\ref{an1}, for this alternative definition of $D(\zeta_A)$ (or of $D(\widetilde{\zeta}_A)$), with $A\subseteq\eR^N$ bounded (as in the present situation), it would suffice to restrict oneself to $\alpha\geq 0$ in the right-hand side of~\eqref{41/2}; this follows since $D(\zeta_A)=\ov{\dim}_BA\geq 0$ and, $D(\zeta_A)=D(\widetilde{\zeta}_A)$ by Remark~\ref{2.71/2} below.
An analogous comment would {\em not} be correct, however, in the case of relative fractal drums discussed in Section~\ref{relative} below.
Indeed, in that case (and with the notation of Section~\ref{relative}), we still have $D:=D(\zeta_A(\,\cdot\,,\O))=\ov{\dim}_B(A,\O)$ (and $D(\zeta_A(\,\cdot\,,\O))=D(\widetilde{\zeta}_A(\,\cdot\,,\O))$), but we may have $\ov{\dim}_B(A,\O)\leq 0$ and even $\ov{\dim}_B(A,\O)=-\ty$; see Proposition~\ref{ndim} and Corollary~\ref{flat} below.}
\begin{equation}\label{41/2}
D(f)=\inf\left\{\alpha\in\eR\,:\,\int_E\varphi(x)^{\alpha}\D|\mu|(x)<\ty\right\},
\end{equation}
where (as above) $|\mu|$ is the total variation measure of $\mu$.

\medskip

The following corollary of Theorem~\ref{an} shows that the distance zeta function can serve as a new tool for computing the upper box dimension of fractal sets in Euclidean spaces.\footnote{More information about other equivalent forms for the computation of the upper box dimension of sets can be found in \cite{laprozu}.}

\medskip

\begin{cor}[\cite{fzf}]\label{an1}
If $A$ is any bounded subset of $\eR^N$, then
\begin{equation}\label{ddim}
\ov\dim_BA=D(\zeta_A).
\end{equation}
Furthermore, in general, we have
\begin{equation}\label{51/4}
-\ty\leq D_{hol}(\zeta_A)\leq D(\zeta_A)=\ov{\dim}_BA.
\end{equation}
\end{cor}

\medskip

\begin{cor}[\cite{fzf}]\label{2.31/2}
Let $A$ be a bounded subset of $\eR^N$ which satisfies the hypotheses of part $(c)$ of Theorem~\ref{an}.
Then, we have
\begin{equation}\label{51/2}
{\dim}_BA=D(\zeta_A)=D_{hol}(\zeta_A).
\end{equation}
\end{cor}

\medskip

\begin{remark}\label{2.41/2}
For most examples of fractal subsets of $\eR^N$ considered in [LapRa\v Zu1--7], 
we have $D(\zeta_A)=D_{hol}(\zeta_A)$.
On the other hand, we leave it to the interested reader to check that in the trivial example of the unit interval $I:=[0,1]\subset\eR$, we have a strict inequality in Corollary~\ref{an1}; more specifically, we have $D_{hol}(\zeta_I)=0<1=D(\zeta_I)=\dim_BI$.\footnote{Note that one has to interpret $0^{\sigma}$ by $+\ty$ for $\sigma<0$ in the integral occurring in Equation~\eqref{z}, which is natural.}
At this point, we do not know whether the second inequality in~\eqref{51/4} is sharp if $D<N$; in other words, it is an open problem to find (if possible) an example satisfying the strict inequalities $D_{hol}(\zeta_A)<D(\zeta_A)<N$.
\end{remark}

\medskip

For a given bounded set $A$, it is of interest to know the corresponding poles of the associated distance zeta function $\zeta_A$,  meromorphically extended (if possible) to a neighborhood of the critical line. Following the terminology of \cite{lapidusfrank12}, these poles are called the {\em complex dimensions} of $A$. We pay particular attention to the set of complex dimensions of $A$ located on the critical line $\{\re s=D(\zeta_A)\}$, which we call  the set of {\em principal complex dimensions of $A$} and denote by $\dim_{PC} A$.

\medskip

For example, it is well known that for the ternary Cantor set $C^{(1/3)}$ we have that $\dim_BC^{(1/3)}=\log_32$ and (see \cite{lapidusfrank12})
$$
\dim_{PC}C^{(1/3)}= \log_32+\frac{2\pi\mathrm{i}}{\log3}\Ze.
$$

\medskip

\begin{xtolerant}{500}{}
The following result provides an interesting connection between the residue of the distance zeta function of a fractal set and its Minkowski contents.
\end{xtolerant}

\medskip

\begin{theorem}[\cite{fzf}]\label{pole1}
Assume that $A$ is a bounded subset of $\eR^N$ which is nondegenerate 
$($that is, $0<\M_*^D(A)\le\M^{*D}(A)<\ty$ and hence, $\dim_BA=D)$, 
and $D<N$. If $\zeta_A(s)=\zeta_A(s,A_\delta)$ can be extended meromorphically to a neighborhood of $s= D$,
then $D$ is necessarily a simple pole of $\zeta_A(s,A_\delta)$, and 
\begin{equation}\label{res}
(N-D)\M_*^D(A)\le\res(\zeta_A(\,\cdot\,,A_\delta),D)\le(N-D)\M^{*D}(A).
\end{equation}
 Furthermore, the value of $\res(\zeta_A(\,\cdot\,,A_\delta), D)$ does not depend on $\delta>0$.
In particular, if $A$ is Minkowski measurable, then 
\begin{equation}\label{pole1minkg1=}
\res(\zeta_A(\,\cdot\,,A_\delta), D)=(N-D)\M^D(A).
\end{equation}
\end{theorem}

\medskip

The distance zeta function is closely related to the tube zeta function of a fractal set.
The latter (fractal) zeta function is defined via the tube function $t\mapsto|A_t|$, $t>0$, of the fractal set $A$, as we now explain.

\medskip

\begin{defn}[\cite{fzf}]\label{zeta_tilde}  
Let $A$ be a bounded set in $\eR^N$. Then, for a given $\d>0$, the {\em tube zeta 
function}
 of $A$, denoted by $\widetilde\zeta_A$, is defined by
\begin{equation}\label{tildz}
\widetilde\zeta_A(s)=\int_0^\delta t^{s-N-1}|A_t|\,\D t,
\end{equation}
for all $s\in\Ce$ with $\re s$ sufficiently large.
\end{defn} 

\medskip

It is shown in [LapRa\v Zu1--4] 
that the distance and tube zeta functions associated with a fractal set $A$ are connected as follows:\footnote{We write here $\zeta_A(\,\cdot\,,A_{\delta}):=\zeta_A$ and $\widetilde{\zeta}_A(\,\cdot\,,A_{\delta}):=\widetilde{\zeta}_A$, for emphasis.}
\begin{equation}\label{equ_tilde}
\zeta_A(s,A_\delta)=\delta^{s-N}|A_\delta|+(N-s)\widetilde\zeta_A(s,A_\delta),
\end{equation}
for any $\delta>0$ and for all $s$ such that $\re s>\ov\dim_BA$.\footnote{In light of the principle of analytic continuation and Remark~\ref{2.71/2}, one deduces that identity~\eqref{equ_tilde} continues to hold whenever one (and hence, both) of the fractal zeta functions $\zeta_A$ and $\widetilde\zeta_A$ is meromorphic on a given domain $U\subseteq\Ce$.}
Using this result, it is easy to reformulate Theorem \ref{pole1} in terms of the tube zeta functions.
In particular, we conclude that the residue of the tube zeta function of a fractal set, computed at $s=D$, is equal to
 its Minkowski content, provided the set is Minkowski measurable.
Moreover, it can be shown directly that the condition $D<N$ from Theorem~\ref{pole1} can be removed in this case.

\medskip

\begin{theorem}[\cite{fzf}]\label{pole1mink_tilde}
Assume that $A$ is a nondegenerate bounded set in $\eR^N$ $($so that $D:=\dim_BA$ exists$)$, 
and there exists a meromorphic extension of $\widetilde\zeta_A$ to a neighborhood of $D$. Then,  $D$ is a simple  pole,
and for any positive $\delta$, $\res(\widetilde{\zeta}_A,D)$ is independent of $\delta$. Furthermore, we have
\begin{equation}\label{zeta_tilde_M}
\M_*^D(A)\le\res(\widetilde\zeta_A, D)\le \M^{*D}(A).
\end{equation}
In particular, if $A$ is Minkowski measurable, then 
\begin{equation}\label{zeta_tilde_Mm}
\res(\widetilde\zeta_A, D)=\M^D(A).
\end{equation}
\end{theorem}

\medskip

A class of fractal sets $A$ for which we have strict inequalities in (\ref{zeta_tilde_M}) (and hence also in~\eqref{res} of Theorem~\ref{pole1} above)\footnote{See relation~\eqref{133/4} of Remark~\ref{2.71/2}.} is constructed in
Theorem \ref{nonmeasurable}; see (\ref{res_inequalities}) below.

\medskip

\begin{remark}\label{2.71/2}
For the purpose of part $(a)$--$(d)$ of this remark, we let $D:=\ov{\dim}_BA$, where $A$ is a bounded subset of $\eR^N$.\footnote{In part $(e)$ of this remark, we let $D:=\ov{\dim}_B(A,\O)$, the relative upper Minkowski dimension of the relative fractal drum $(A,\O)$, defined as in Equation~\eqref{dimrel} of Section~\ref{relative} below.}

\medskip

$(a)$~In light of~\eqref{equ_tilde}, it is immediate to see that provided $D<N$ (or, equivalently, $D\neq N$ since we always have $D\leq N$), the distance and tube zeta functions contain essentially the same information.
More specifically, given any domain (i.e., connected open subset) $U$ of $\Ce$, $\zeta_A$ and $\widetilde{\zeta}_A$ can either be simultaneously meromorphically extended to $U$ or else, neither of them can be meromorphically extended to $U$.
In the former case, by analytic continuation, the functional equation~\eqref{equ_tilde} continues to hold throughout $U$.
In particular, the (unique) meromorphic extensions of $\zeta_A$ and $\widetilde{\zeta}_A$ have the same poles in $U$ (i.e., they have the same {\em visible complex dimensions in $U$}) and (in the case of simple poles) their respective residues at $\omega\in U$ are simply related as follows:
\begin{equation}\label{131/2}
\res({\zeta_A,\omega})=(N-\omega)\res({\widetilde{\zeta}_A,\omega}).
\end{equation}
In particular, if $D \in U$, then we have
\begin{equation}\label{133/4}
\res({\zeta_A,D})=(N-D)\res({\widetilde{\zeta}_A,D}).
\end{equation}
This last relation,~\eqref{133/4}, combined with the above discussion, helps explain how to go from Theorem~\ref{pole1}  to Theorem~\ref{pole1mink_tilde}  (and, in particular, from~\eqref{pole1minkg1=} to~\eqref{zeta_tilde_Mm} above).
In the special case when $D=N$, we have a similar conclusion as above, except that in this case, if $N\in U$, then $s=N$ is a simple pole of the tube zeta function $\widetilde{\zeta}_A$, whereas it is a removable singularity of the distance zeta function ${\zeta}_A$.
More precisely, we always have
\begin{equation}
\zeta_A(N,\d)=|A_\d|-\res(\widetilde{\zeta}_A,N).
\end{equation}
Furthermore, if $\dim_BA=N$ and if $A$ is Minkowski measurable, then by Theorem~\ref{pole1mink_tilde}, we have 
\begin{equation}
\zeta_A(N,\d)=|A_\d|-\mathcal{M}^N(A)=|A_\d|-|\ov{A}|.
\end{equation}

\medskip

(b)~It follows from the above discussion (in part $(a)$ of this remark) and the relevant definitions previously given in this section that, still for $D:=\ov{\dim}_BA$, we have
\begin{equation}\label{134/5}
D=\ov{\dim}_BA=D(\zeta_A)=D(\widetilde{\zeta}_A),
\end{equation}
and hence, that $\zeta_A$ and $\widetilde{\zeta}_A$ have the same half-plane of (absolute) convergence,
\begin{equation}\label{135/6}
\{\re s>D\}=\Pi(\zeta_A)=\Pi(\widetilde{\zeta}_A),
\end{equation}
as well as the same critical line, $\{\re s=D\}$, with $D$ given by~\eqref{134/5} above.\footnote{Clearly, $\zeta_A$ and $\widetilde{\zeta}_A$ also have the same abscissae of meromorphic convergence, $D_{mer}(\zeta_A)=D_{mer}(\widetilde{\zeta}_A)$, and hence, the same half-plane of meromorphic convergence, defined in exactly the same way as their counterparts for holomorphic convergence, except for ``holomorphic'' replaced by ``meromorphic''.}
Furthermore, if $D<N$, then $D_{hol}(\zeta_A)=D_{hol}(\widetilde{\zeta}_A)$, and hence, $\zeta_A$ and $\widetilde{\zeta}_A$ also have
the same half-plane of holomorphic convergence, $\mathcal{H}(\zeta_A)=\mathcal{H}(\widetilde{\zeta}_A)$. 

\medskip

$(c)$ We note that Theorem~\ref{an} and its corollaries, Corollaries~\ref{an1} and~\ref{2.31/2}, have an exact counterpart for tube zeta functions. 
Namely, it suffices to replace $\zeta_A$ by $\widetilde{\zeta}_A$ in the statement of these results, and moreover, one does not need the condition $D<N$ assumed in Theorem \ref{an}$(c)$ and Corollary \ref{2.31/2}.
Analogous comments could be made about all the relevant results discussed in this paper (for example, Theorems~\ref{measurable} and~\ref{nonmeasurable} in Section~\ref{merom_ext} or Proposition~\ref{Cmap} along with Theorems~\ref{trans} and~\ref{quasi1} in Section~\ref{qp_sets} below).
It suffices to take into account the relations between the residues of $\zeta_A$ and $\widetilde{\zeta}_A$ (as given in~\eqref{131/2} or, in particular, in~\eqref{133/4} above).

\medskip

$(d)$~Many other illustrations of the direct and simple connections between the fractal zeta functions $\zeta_A$ and $\widetilde{\zeta}_A$, as well as of the use of these zeta functions in a given situation or example, are provided throughout~\cite{fzf} (as well as in [LapRa\v Zu2--7]).

\medskip

$(e)$~An entirely analogous comment can be made in the more general situation of a relative fractal drum $(A,\O)$ discussed in Section~\ref{relative} below, and the associated (relative) distance and tube zeta functions, $\zeta_A(\,\cdot\,,\O)$ and $\widetilde{\zeta}_A(\,\cdot\,,\O)$; see~\cite[Chapter~4]{fzf} along with~[LapRa\v Zu2--7].
\end{remark}

\section{Meromorphic extensions of fractal zeta functions}\label{merom_ext}

We begin this section by recalling some basic facts, terminology and notation from~\cite{lapidusfrank12} and~\cite{fzf}.

\medskip

Following [Lap--vFr1--3] 
and [LapRa\v Zu1--7], 
where the extension to higher dimensions of the theory of complex dimensions is developed, we adopt the following terminology and notation.
Given a domain $U$ of $\Ce$ to which a fractal zeta function $f=f(s)$ (attached to a given bounded set $A\subseteq\eR^N$) can be meromorphically extended (necessarily uniquely, according to the principle of analytic continuation), we call the poles (of the meromorphic continuation) of $f$ in $U$ the {\em visible complex dimensions of} $A$, or simply, the {\em complex dimensions} (if no ambiguity may arise or else if, as is often the case in many important examples, if $U=\Ce$).
Correspondingly, we denote by $\po(f)$ or, if necessary, to avoid ambiguities, by $\po(f,U)$, the {\em set of all} (visible) {\em complex dimensions} (in $U$).
In particular, if $U=\Ce$, we simply write $\po(f)$ instead of $\po(f,\Ce)$.

\medskip

In light of Remark~\ref{2.71/2}, provided $D<N$, the set $\po(f,U)=\po(f)$ of (visible) complex dimensions of $A$ is the same, whether $f:=\zeta_A$ or $f:=\widetilde{\zeta}_A$, and (in the case of simple poles) the associated residues at a given pole $\omega\in U$ are related by~\eqref{131/2} (and, in particular, by~\eqref{133/4} in the important case where $\omega:=D$, the upper box dimension of $A$).\footnote{More generally, in the case of multiple poles, there is a similar relation connecting the principal parts of $\zeta_A$ and $\widetilde{\zeta}_A$, in light of the functional equation~\eqref{equ_tilde}.}
Therefore, it is justified to also refer to $\po(f)$ as the {\em set of $($visible$)$ complex dimensions of} $A$.
Furthermore, when the domain $U$ contains $\{\re s\geq D\}$, then $\dim_{PC}A$, the set of complex dimensions of $A$ located on the {\em critical line} $\{\re s=D\}$, is called the {\em set of principal complex dimensions of} $A$.
It is clearly independent of the choice of the domain $U\subseteq\Ce$ as above.\footnote{In light of Theorem~\ref{an} and Remark~\ref{2.71/2}, and provided $D<N$, $\dim_{PC}A$ is also independent of the choice of either $\zeta_A$ or $\widetilde{\zeta}_A$.
Furthermore, since $\zeta_A$ and $\widetilde{\zeta}_A$ are holomorphic on the open half-plane $\{\re s>D\}$, $\dim_{PC}A$ then consists of the complex dimensions of $A$ {\em with the largest possible real part},
$\re s=D$, where $D=\ov{\dim}_BA=D(\zeta_A)=D(\widetilde{\zeta}_A)$.}

\medskip

Moreover, following~[LapRa\v Zu1--7], 
in Section~\ref{relative} below, an entirely analogous definition and terminology will be adopted for the (relative) fractal zeta functions $\zeta_A(\,\cdot\,,\O)$ and $\widetilde{\zeta}_A(\,\cdot\,,\O)$ of relative fractal drums $(A,\O)$ (instead of the more special case of bounded subsets of $\eR^N$).

\medskip

Finally, following~[Lap--vFr1--3], 
we will talk in a similar manner about the set $\po(\zeta_{\mathcal L})=\po(\zeta_{\mathcal L},U)$ of (visible) complex dimensions of a fractal string $\mathcal L$, that is, of the (visible) poles of the geometric zeta function of $\mathcal L$.
(For the appropriate definitions and notation see~\cite[Chapter~1]{lapidusfrank12}.)

\medskip

\begin{remark}\label{footnote}
According to some of the results obtained in~\cite[Chapter~2]{fzf} (briefly discussed in the text surrounding Equation~\eqref{dist_geo} above), and provided $D<1$ and $U\subseteq\{\re s>0\}$ (or, more generally, $U\subseteq\Ce\setminus\{0\}$), then $\po(\zeta_{\mathcal L})$ coincides with both $\po(\zeta_A)$ and $\po(\widetilde{\zeta}_A)$, where $A:=\partial\mathcal L$ denotes the boundary of the fractal string $\mathcal L$.
For the same reason, provided $0<D<1$ (note that we always have $D=\ov{\dim}_B\partial\mathcal L\in[0,1]$), $\dim_{PC}\partial\mathcal L$ can be defined either via $\zeta_{\mathcal L}$, $\zeta_A$ or $\widetilde{\zeta}_A$, and independently of the resulting choice as well as of the geometric realization of $\L$.
\end{remark}

\medskip

\begin{xtolerant}{1000}{}
In particular, recall that a fractal string $\mathcal L$ is viewed as a bounded open subset $\O$ of $\eR$.
Consider the sequence of lengths $(l_j)_{j=1}^{\ty}$, written in nonincreasing order and written according to multiplicity, of the connected components (i.e., open intervals) of $\O$.
The {\em geometric zeta function} $\zeta_{\mathcal L}$ of $\mathcal L$ is then defined (for all $s\in\Ce$ with $\re s$ sufficiently large) by
\begin{equation}\label{131/6}
\zeta_{\mathcal L}(s):=\sum_{j=1}^{\ty}l_j^s.
\end{equation}
Its abscissa of convergence, $D(\zeta_{\mathcal L})$, coincides with the inner (upper) Minkowski (or box) dimension of $\partial\mathcal L$, the boundary of $\mathcal L$, defined as $\partial\O$ (see~[Lap2--3] 
and~\cite[Theorem~1.10,~p.~17]{lapidusfrank12}):\footnote{Here and thereafter, we assume implicitly that $\mathcal L$ is nontrivial; that is, the sequence $(l_j)_{j=1}^{\ty}$ is not finite (equivalently, $\O$ consists of infinitely many intervals) and hence, $l_j\searrow 0$.}
\begin{equation}\label{131/5}
D:=D(\zeta_{\mathcal L})=\ov{\dim}_B\partial\mathcal L.
\end{equation}
In particular, it is independent of the representation of $\mathcal L$ as a bounded open subset of $\eR$ and depends only on the sequence of lengths $(l_j)_{j=1}^{\ty}$ defining $\mathcal L$.
This observation, first made in~[Lap2--3], 
has played a crucial role in the development of the theory of complex dimensions of fractal strings developed in~[Lap--vFr1--3] 
(and now, of its higher-dimensional counterpart in~[LapRa\v Zu1--7] 
as well as in~\cite{laprozu} which provides another higher-dimensional counterpart to this theory).
It also follows from the results of~[Lap2--3] 
(and of a well-known fact concerning generalized Dirichlet series with positive coefficients; see, e.g.,~\cite{serre} or~\cite{Pos}) that $D$ also coincides with the abscissa of holomorphic convergence of $\zeta_{\mathcal L}$; so that, in light of~\eqref{131/5}, we have
\begin{equation}\label{131/5+e}
D:=D(\zeta_{\mathcal L})=D_{hol}(\zeta_{\mathcal L})=\ov{\dim}_B\partial\mathcal L,
\end{equation}
for any (nontrivial) fractal string $\mathcal L$, independently of its geometric realization as a bounded open set $\O\subseteq\eR$ and only depending on the sequence of lengths (or, more generally, `scales') $\mathcal L=(l_j)_{j=1}^{\ty}$.
(See~\cite{lapidusfrank12}.)
As was recalled in Remark~\ref{footnote}, the same is true for the set of (visible) complex dimensions of $\L$ (provided $U\subseteq\Ce\setminus\{0\}$) and the set of principal complex dimensions of $\L$.
\end{xtolerant}

\medskip

The following comment may be helpful to the reader in placing in a proper context several results discussed in this paper.
(See, for example, Theorem~\ref{pole1}, Theorem~\ref{pole1mink_tilde}, Problem~\ref{8.7}, Problem~\ref{8.8} and Remark~\ref{8.81/2}.)

\medskip

\begin{remark}\label{3.11/2}
The complex dimensions of a variety of fractal strings are calculated in~[Lap--vFr1--3].
In particular, for self-similar strings, the structure of the complex dimensions is precisely determined in~\cite[Chapters~2 and~3]{lapidusfrank12}.
In the lattice case, the complex dimensions are periodically distributed along finitely many vertical lines (all with abscissa not exceeding $D$ and with the same oscillatory period), including on the vertical line $\{\re s=D\}$, whereas in the nonlattice case, they are quasiperiodically distributed and (in contrast to the lattice case) $D$ is the only complex dimension with real part $D$ (while there exists a sequence of complex dimensions tending to, but not touching, the vertical line $\{\re s=D\}$).\footnote{The lattice/nonlattice dichotomy can be briefly defined as follows, for a self-similar string $\mathcal{L}$ of scaling ratios $0<r_1,\ldots,r_q<1$.
If the distinct values of the scaling ratios are all integer powers of a same number $r\in(0,1)$, then $\mathcal{L}$ is said to be {\em lattice}, while it is said to be {\em nonlattice}, otherwise.
The {\em oscillatory period} of a lattice string is then given by $\mathbf{p}:=2\pi/\log(1/r)$, where $r$ is the smallest such number.}
In every case, $D$ is the only complex dimension with real part $D$ and it is a simple pole of $\zeta_{\mathcal{L}}$.
Moreover, it is shown in~\cite[Chapter~3]{lapidusfrank12} that a given nonlattice string can be approximated by a sequence of lattice strings with larger and larger oscillatory periods, and analogously, for the associated complex dimensions.
(Many additional results about the specific ``quasiperiodic'' structure of the complex dimensions of nonlattice strings can be found in {\em loc.\ cit.})
Finally, it is shown in~\cite[Section~8.4]{lapidusfrank12} that a self-similar string $\mathcal{L}$ is always Minkowski nondegenerate, and is Minkowski measurable if and only if $D$ is the only complex dimension located on the critical line $\{\re s=D\}$ (and it is simple), that is, if and only if $\mathcal{L}$ is a nonlattice string.
In that case, the Minkowski content of $\mathcal{L}$ is given by $\mathcal{M}={2^{1-D}}\res(\zeta_{\mathcal{L}},D)/{D(1-D)}$; compare with Theorem~\ref{pole1}, Theorem~\ref{pole1mink_tilde} and Remark~\ref{2.71/2}$(a)$, in particular.
Some of these results have been extended to higher dimensions for certain self-similar tilings and a special class of self-similar sets, in~[LapPe1--2] 
and later, more generally, in~[LapPeWi1--2].
\end{remark}

\medskip

\begin{remark}\label{31/2}
Fractal strings were introduced in~[LaPo1--2], 
building on examples studied in~\cite{Lap1}.
They were used in a variety of situations, including to obtain a geometric reformulation of the Riemann hypothesis in terms of inverse spectral problems ([LapMa1--2], 
revisited and extended from various points of view in~\cite{lapidushe}, [Lap--vFr1--3] 
and, more recently, in [HerLap1--5] 
and~[Lap6--10]), 
to develop the one-dimensional theory of complex dimensions [Lap--vFr1--3], 
to explore aspects of $p$-adic analysis~\cite{LaLu} (as described in~\cite[Section~13.2]{lapidusfrank12}), analysis on fractals ([Lap3--5], 
\cite{ChrIvLa}, [Tep1--2], 
[LalLap1--2], 
\cite{LaSar}) and multifractals (\cite{lapidusrock}, \cite{LaLeRo}, \cite{lemen}, \cite{ElLaMaRo}, as described in part in~\cite[Section~13.3]{lapidusfrank12}), random fractal strings~\cite{HamLa} (as described in~\cite[Section~13.4]{lapidusfrank12}), as well as to develop a higher-dimensional theory of complex dimensions in the important, but still very special case of `fractal sprays' (as introduced in~\cite{LapPo2} and further studied in [Lap2--3], 
[Lap--vFr1--3] 
and [LapPe2--3], 
\cite{pe2}, \cite{pewi}, [LapPeWi1--2], 
as described, in particular, in~\cite[Section~13.1]{lapidusfrank12}) and a quantized analog of fractal strings, called `fractal membranes' (\cite{lapz}, as briefly described in~\cite[Section~13.5]{lapidusfrank12}). 
\end{remark}

\medskip

We continue the main part of this section by discussing a simple but interesting example, in which we show that the tube zeta function (and hence also the distance zeta function, since $\dim_BA=N-1<N$) of the $(N-1)$-dimensional sphere $A$ in $\eR^N$ can be meromorphically extended to all of $\Ce$, and we calculate the corresponding complex dimensions of $A$ (which are all located on the real axis, as expected since $A$ itself is not fractal).\footnote{In the sense of an extended version of the notion of fractality defined in~[Lap1,Lap--vFr1--3]. 
}

\medskip

\begin{example}\label{sphere} Let $B_R(0)$ be the ball of $\eR^N$ centered at the origin with radius $R>0$, and let $A:=\pa B_R(0)$ be its boundary, i.e., the $(N-1)$-dimensional sphere of radius $R$. 
Then, the tube zeta function can be explicitly computed and one finds that
\begin{equation}\label{131/4}
\begin{aligned}
\widetilde\zeta_A(s)&=\o_N\sum_{k=0}^N(1-(-1)^k)R^{N-k}\binom Nk\frac{\d^{s-N+k}}{s-(N-k)},
\end{aligned}
\end{equation}
for all $s\in\Ce$ with $\re s>N-1$.\footnote{Here, the numbers $\binom Nk$ stand for the usual binomial coefficients.}
Here, $\o_N$ is the $N$-dimensional Lebesgue measure (or volume) of the unit ball of $\eR^N$.\footnote{Hence, $\o_N=\pi^{N/2}/(N/2)!$, where $x!:=\Gamma(x+1)$ and with $\Gamma$ denoting the classic gamma function; so that $x!$ is is the usual factorial function when $x\in\eN$.}
By the principle of analytic continuation, the above expression in~\eqref{131/4} is in fact the meromorphic extension of the tube zeta function to the whole complex plane, and we still denote it by $\widetilde\zeta_A(s)$. It follows that $\dim_BA$ exists and
\begin{equation}\label{dimBAN-1}
\dim_BA=D(\widetilde\zeta_A)=N-1
\end{equation}
and moreover, the set of complex dimensions of $A$ (i.e., the set of poles of $\widetilde\zeta_A$ or, equivalently, of $\zeta_A$), is given by (with $\lfloor x \rfloor$ denoting the integer part of $x\in\eR$)
\begin{equation}\label{dimSN-1}
\begin{aligned}
\po(\widetilde\zeta_A)&=\left\{N-(2j+1):j=0,1,2,\dots,\left\lfloor\frac{N-1}2\right\rfloor\right\}\\
&=\left\{N-1,N-3,\dots,N-\left(2\left\lfloor\frac{N-1}2\right\rfloor+1\right)\right\}.
\end{aligned}
\end{equation}
For an odd $N$, the last number (on the right) in this set is equal to $0$, while for an even $N$, it is equal to $1$.
Furthermore, the residue of the tube zeta function $\widetilde\zeta_A$ at any of its poles $N-k\in\po(\widetilde\zeta_A)$ is equal to
\begin{equation}\label{resSN-1}
\res(\widetilde\zeta_A,N-k)=2\o_N\binom Nk R^{N-k}.
\end{equation}
Since $\binom Nk=\binom N{N-k}$, we can write this result in an even more `symmetric' form: 
\begin{equation}\label{resSN-1b}
\res(\widetilde\zeta_A,m)=2\o_N\binom Nm R^m,\q\mbox{for all\q $m\in\po(\widetilde\zeta_A)$}.
\end{equation}
Clearly, in light of~\eqref{dimBAN-1} and~\eqref{dimSN-1}, we have $\dim_{PC}A=\{N-1\}$  and according to~\eqref{resSN-1} or~\eqref{resSN-1b}, we have
\begin{equation}\label{271/4}
\res(\widetilde{\zeta}_A,N-1)=2N\o_NR^{N-1}.
\end{equation}
Moreover, in light of Remark~\ref{2.71/2}, $\po(\zeta_A)=\po(\widetilde{\zeta}_A)$ is still given by~\eqref{dimSN-1}, and the counterpart for $\zeta_A$ of~\eqref{resSN-1b} and~\eqref{271/4} becomes, respectively,
\begin{equation}\label{271/2}
\res(\zeta_A,m)=2(N-m)\o_N\binom Nm R^m,
\end{equation}
 for all $m\in\po(\zeta_A)$ and
\begin{equation}\label{273/4}
\res(\zeta_A,N-1)=2N\o_NR^{N-1};
\end{equation}
see~\eqref{131/2} or~\eqref{133/4}, respectively.
This concludes Example~\ref{sphere}.
\end{example}

\medskip

Since the definition of the set of complex dimensions of $A$ requires  the existence of a nontrivial meromorphic extension of the distance zeta function $\zeta_A$
to be satisfied, it is natural to study this issue in more detail.

\medskip

The following theorem shows that if we perturb the classical 
Riemann zeta function (see, e.g.,~\cite{Titch2})
\begin{equation}\label{riemann_zeta}
\zeta_R(s)=\sum_{j=1}^\ty j^{-s}
\end{equation}
by a sufficiently small sequence of real numbers $(c_j)_{j\ge1}$, in the sense that $c_j=O(j^{\beta})$ as $j\to\ty$, where $\beta<1$, then the resulting perturbed Riemann zeta function 
\begin{equation}\label{perturbed}
\zeta_{R,pert}(s)=\sum_j(j+c_j)^{-s}
\end{equation} 
possesses a (necessarily unique) meromorphic extension to $\{\re s>\beta\}$. We denote by $D(\zeta_{R,pert})$ the 
abscissa of convergence of $\zeta_{R,pert}(s)$.

\medskip

\begin{theorem}[\cite{fzf}]\label{riemann}
Let $\beta\in(-\ty,1)$ be fixed, and assume that $c_j=O(j^\beta)$ as $j\to\ty$. Then, for the perturbed Riemann zeta function defined by $(\ref{perturbed})$,
we have $D(\zeta_{R,pert})=1$, and $\zeta_{R,pert}$ has a $($necessarily unique$)$ 
meromorphic extension {\rm({\it at least})} to the open half-plane
\begin{equation}\label{beta}
\{\re s>\beta\}.
\end{equation}
Furthermore, $s=1$ is a pole of the meromorphic continuation in this half-plane; it is simple, and $\res(\zeta_{R,pert},1)=1$. The sets of poles of the classical Riemann zeta function and of $\zeta_{R,pert}$, located in $\{\re s>\beta\}$, coincide, which means in the present case that $s=1$ is the only pole of $\zeta_{R,pert}$ in  $\{\re s>\beta\}$.
\end{theorem}

\medskip

An analogous result can be obtained in the context of tube zeta functions of bounded fractal sets. We deal with the case of Minkowski measurable sets first.

\medskip

\begin{theorem}[Minkowski measurable case, \cite{fzf}]\label{measurable}%
\begin{xtolerant}{400}{}
Let $A$ be a bounded subset of $\eR^N$ such that there exist  $\alpha>0$, $\mathcal {M}\in(0,+\ty)$ and $D\ge0$ satisfying
\begin{equation}\label{A_t}
|A_t|= t^{N-D}\left({\mathcal {M}}+O(t^\alpha)\right)\quad\mathrm{as}\quad t\to0^+.
\end{equation}
Then, $\dim_BA$ exists and $\dim_BA=D$. Furthermore, $A$ is Minkowski measurable with Minkowski content $\mathcal {M}^D(A)=\mathcal {M}$. 
Moreover, the tube zeta function $\widetilde\zeta_A$ has for abscissa of convergence $D(\widetilde\zeta_A)=\dim_BA=D$ and possesses a unique meromorphic continuation $($still denoted by $\widetilde\zeta_A)$ to $($at least$)$ the open half-plane $\{\re s>D-\alpha\}$.
The only pole of $\widetilde\zeta_A$ in this half-plane is $s=D$; it is simple, and $\res(\widetilde\zeta_A,D)=\M$.
\end{xtolerant}
\end{theorem}

\medskip

Next, we deal with a class of Minkowski nonmeasurable sets. 
Before stating Theorem \ref{nonmeasurable}, let us first introduce some notation.
Given a $T$-periodic function $G:\eR\to\eR$, we denote by $G_0$ its truncation to $[0,T]$, while the Fourier transform of $G_0$
is denoted by $\hat G_0$:
\begin{equation}\label{fourier}
\begin{aligned}
G_0(\tau)&=
\begin{cases}
G(t)& \mbox{if $\tau\in[0,T]$}\\
0& \mbox{if $\tau\notin[0,T]$},
\end{cases}\\
\hat G_0(t)&=\int_{-\ty}^{\ty}{\E}^{-2\pi {\I}t\tau}G_0(\tau)\,\D\tau=\int_0^T{\E}^{-2\pi {\I}t\tau}G(\tau)\,\D\tau.
\end{aligned}
\end{equation}

\medskip

\begin{theorem}[Minkowski nonmeasurable case; \cite{fzf}]\label{nonmeasurable}%
Let $A$ be a bounded subset of $\eR^N$ such that there exist $D\ge0$, $\alpha>0$, and $G:\eR\to(0,+\ty)$ a nonconstant periodic function with period $T>0$, 
satisfying
\begin{equation}\label{G}
|A_t|=t^{N-D}\left(G(\log t^{-1})+O(t^\alpha)\right)\quad\mbox{as\qs$t\to0^+$.}
\end{equation}
 Then $G$ is continuous, $\dim_BA$ exists and $\dim_BA=D$. Furthermore, $A$ is Minkowski nondegenerate
 with upper and lower Minkowski contents respectively given by
\begin{equation}\label{1.4.201/2}
\M_*^D(A)=\min G,\quad \M^{*D}(A)=\max G.
\end{equation}
Moreover, the tube zeta function $\widetilde\zeta_A$ has for abscissa of convergence $D(\widetilde\zeta_A)=D$ and possesses a unique meromorphic extension $($still denoted by $\widetilde\zeta_A$$)$
to {\rm({\it at least})} the half-plane $\{\re s>D-\alpha\}$.

\medskip

In addition, the set of all the poles of $\widetilde\zeta_A$ located in this half-plane is given by\footnote{Note that the set defined by (\ref{Dpoles})
coincides with the set of principal complex dimensions of $A$, that is, with $\dim_{PC}A$.}
\begin{equation}\label{Dpoles}
\mathcal {P}(\widetilde \zeta_A)=\left\{s_k=D+\frac{2\pi}T{\I}k:\hat G_0(\frac kT)\ne0,\,\,k\in\Ze\right\}
\end{equation}
{\rm({\it see} (\ref{fourier}))}; they are all simple, and the residue at each $s_k\in\mathcal {P}(\widetilde\zeta_A)$, $k\in\Ze$, is given by
\begin{equation}\label{res_fourier}
\res(\widetilde\zeta_A,s_k)=\frac1T\hat G_0(\frac kT).
\end{equation}
If $s_k\in \mathcal {P}(\widetilde \zeta_A)$, then $s_{-k}\in \mathcal {P}(\widetilde \zeta_A)$ $($in agreement with the `reality principle'$)$, and
\begin{equation}\label{riemann_lebesgue}
|\res(\widetilde\zeta_A,s_k)|\le \frac1T\int_0^TG(\tau)\,\D\tau; 
\end{equation}
furthermore, $\lim_{k\to\pm\ty}\res(\widetilde\zeta_A,s_k)=0$.

\medskip

Moreover, the set of poles $\mathcal {P}(\widetilde\zeta_A)$ $($i.e., of complex dimensions of $A)$ contains $s_0=D$, and
\begin{equation}\label{avarage}
\res(\widetilde\zeta_A,D)=\frac1T\int_0^TG(\tau)\,\D\tau.
\end{equation}
In particular, $A$ is {\rm not} Minkowski measurable and
\begin{equation}\label{res_inequalities}
\M_*^D(A)<\res(\widetilde\zeta_A,D)<\M^{*D}(A).
\end{equation}
\end{theorem}

\medskip

\begin{remark}\label{3.41/2}
In light of Remark~\ref{2.71/2} and if additionally $D<N$, the results of Example~\ref{sphere}, Theorem~\ref{measurable} and Theorem~\ref{nonmeasurable} have an obvious counterpart for the distance zeta function $\zeta_A$ (instead of the tube zeta function $\widetilde{\zeta}_A$).
The statements of the corresponding results are identical, except for the fact that the values of the residues of $\zeta_A$ are expressed in the terms of the corresponding residues of $\widetilde{\zeta}_A$ via the relation~\eqref{131/2} at a simple pole $\omega$ (and, in particular, via the relation~\eqref{133/4} if $\omega=D$).
(We gave a specific illustration of this observation at the end of Example~\ref{sphere}.)
An entirely analogous comment can be made about the results stated and the examples discussed in Sections~\ref{qp_sets} and~\ref{relative} below.
For the most part, we will not do so, however, in order to avoid unnecessary repetitions.
\end{remark}

\begin{remark}\label{ftff}
A partial converse of Theorems~\ref{measurable} and~\ref{nonmeasurable} is provided in [LapRa\v Zu6--7].
More precisely, in \cite{crasext_ftf} (announced in~\cite{cras_ftf}), pointwise and distributional tube formulas for general bounded subsets of $\eR^N$ $(N\geq 1)$ that satisfy a languidity condition\footnote{See~\cite[Section~5.3]{lapidusfrank12} for the definition of this notion.} are derived in terms of a sum over the visible complex dimensions.
(See also Problem~\ref{8.8} and Remark~\ref{8.81/2} at the end of Section~\ref{open} below for additional information.)
Similarly as in~[Lap--vFr1--3] 
where ``fractal tube formulas'' are obtained in the case of fractal strings (see, especially, \cite[Section~8.1]{lapidusfrank12}), the use of the inverse Mellin transform applied to the tube zeta function is the main idea behind the proofs of these results in
~[LapRa\v Zu6--7].\footnote{See, especially, \cite[Chapters~5 and~8]{lapidusfrank12}; see also the case of certain classes of fractal sprays and self-similar tilings later treated (via ``tubular zeta functions'') in [LapPe2--3] 
and [LapPeWi1--2] 
(as described in \cite[Section~13.1]{lapidusfrank12}).}
As a consequence of these formulas and a certain Tauberian theorem due to Wiener and Pitt (which generalizes Ikehara's Tauberian theorem, see~\cite{Pos}), a Minkowski measurability criterion is also given in~[LapRa\v Zu6--7].
It is expressed in terms of nonexistence of nonreal complex dimensions on the critical line $\{\re s=D\}$, and generalizes to higher dimensions the corresponding criterion obtained for fractal strings (i.e., for $N=1$) in~[Lap--vFr1--3]; 
see, especially, \cite[Section~8.3]{lapidusfrank12}.
\end{remark}

\section{Transcendentally quasiperiodic sets}\label{qp_sets}

In this section, we define a class of {\em quasiperiodic fractal sets}. The simplest of such sets has two incommensurable periods. Moreover, using generalized Cantor sets, it is possible to conclude that
the quotient of their periods is a transcendental real number.

\medskip

A construction of such sets is based on a class of generalized Cantor sets with two parameters, which we now introduce.

\medskip

\begin{defn}[\cite{fzf}]\label{Cma}
The generalized Cantor sets $C^{(m,a)}$ are determined by an integer $m\ge2$ and a positive real number $a$ such that $ma<1$.
In the first step of Cantor's construction, we start with $m$ equidistant, closed intervals in $[0,1]$ of length $a$, with $m-1$ `holes', each of length $(1-ma)/(m-1)$. In the second step, we continue by scaling by the factor $a$ each of the $m$ intervals of length $a$, and so on, ad infinitum.
The  $($two-parameter$)$ {\em generalized Cantor set} $C^{(m,a)}$ is defined as the intersection of the decreasing sequence of compact sets constructed in this way.
It is easy to check that $C^{(m,a)}$ is a perfect, uncountable compact subset of $\eR$; furthermore, $C^{(m,a)}$ is also self-similar.
For $m=2$, the sets $C^{(m,a)}$ are denoted by  $C^{(a)}$.
The classic ternary Cantor set is obtained as $C^{(2,1/3)}$.
In order to avoid any possible confusion, we note that the generalized Cantor sets introduced here are different from the generalized Cantor strings introduced and studied in~\cite[Chapter 10]{lapidusfrank12}, as well as used in~\cite[Chapter 11]{lapidusfrank12}.
\end{defn}

\medskip

We collect some of the basic properties of generalized Cantor sets in the following proposition.

\medskip

\begin{prop}[\cite{fzf}]\label{Cmap}
 If $C^{(m,a)}\subset\eR$ is the generalized Cantor set introduced in Definition~\ref{Cma}, 
where $m$ is an integer, $m\ge2$, and $a\in(0,1/m)$,
then\footnote{We mention in passing that the box dimension of $C^{(m,a)}$ is equal to its Hausdorff dimension.
 The proof of this fact in the case of the classical Cantor set can be found in \cite{falc}; see also \cite{hutchinson}.}
\begin{equation}\label{2.1.1}
D:=\dim_B C^{(m,a)}=D(\zeta_A)=\log_{1/a}m.
\end{equation}
Furthermore, the tube formula associated with $C^{(m,a)}$ is given by
\begin{equation}\label{Cmat}
|C^{(m,a)}_t|=t^{1-D}G(\log\frac 1t)
\end{equation}
for $t\in(0,\frac{1-ma}{2(m-1)})$, where $G=G(\tau)$ is the following nonconstant periodic function, with minimal period equal to $T=\log (1/a)$, defined by 
\begin{equation}\label{Gtau}
G(\tau)=c^{D-1}(ma)^{g\left(\frac{\tau-c}{T}\right)}+2\,c^Dm^{g\left(\frac{\tau-c}{T}\right)}.
\end{equation}
Here, $c=\frac{1-ma}{2(m-1)}$ and $g:\eR\to[0,+\ty)$ is the $1$-periodic function defined by $g(x)=1-x$ for every $x\in(0,1]$.

\medskip

Moreover,
\begin{equation}
\begin{aligned}
\mathcal{M}_*^D(C^{(m,a)})&=\frac1D\left(\frac{2D}{1-D}\right)^{1-D},\\
\mathcal{M}^{*D}(C^{(m,a)})&=\left(\frac{1-ma}{2(m-1)}\right)^{D-1}\frac{m(1-a)}{m-1}.
\end{aligned}
\end{equation}

\medskip

Finally, if we assume that $\delta\ge\frac{1-ma}{2(m-1)}$, then the distance zeta function of $A=C^{(m,a)}$ is given by
\begin{equation}\label{zetaCma}
\zeta_A(s)=\left(\frac{1-ma}{2(m-1)}\right)^{s-1}\frac{1-ma}{s(1-ma^s)}+\frac{2\delta^s}s.
\end{equation}
As a result, $\zeta_A(s)$ admits a meromorphic continuation to all of $\Ce$, given by the right-hand side of Equation $(\ref{zetaCma})$. In particular, the set of poles of $\zeta_A(s)$ $($in $\Ce)$ and the residue of $\zeta_A(s)$ at $s=D$ are respectively given by
\begin{equation}\label{2.1.6}
\begin{aligned}
\po(\zeta_A)&=(D+\mathbf p{\I}\Ze)\cup\{0\}\q\mathrm{and}\\ 
\res(\zeta_A,D)&=\frac{1-ma}{DT}\left(\frac{1-ma}{2(m-1)}\right)^{D-1},
\end{aligned}
\end{equation}
where $\mathbf p=2\pi/T$  is the oscillatory period $($in the sense of {\rm\cite{lapidusfrank12}}$)$.
 Finally, each pole in $\po(\zeta_A)$ is simple.
\end{prop}

\medskip

The definition of quasiperiodic sets is based on the following notion of quasiperiodic functions, which will be useful for our purposes.\footnote{We note that part of Definition~\ref{quasip} is based on the definition from \cite{enc} and is very different from the usual definition of Bohr-type quasiperiodic functions.}

\medskip

\begin{defn}[\cite{fzf}]\label{quasip}
We say that a function $G=G(\tau):\eR\to\eR$ is {\em transcen\-dentally 
$n$-quasiperiodic}
 if it is of the form $G(\tau)=H(\tau,\dots,\tau)$,
where $H:\eR^n\to\eR$  is a function that is nonconstant and $T_k$-periodic in its $k$-th component, for each $k=1,\dots,n$, and the periods $T_1,\dots, T_n$ are algebraically (and hence, rationally) independent. The values of $T_i$ are called the 
{\em quasiperiods of $G$}. If the set of quasiperiods $\{T_1,\dots,T_n\}$ is rationally independent and algebraically dependent, we say that {\em $G$ is algebraically $n$-quasiperiodic}.
\end{defn}

\medskip

\begin{defn}[\cite{fzf}]\label{quasiperiodic}
Given a bounded subset $A\st\eR^N$, we say that a function $G:\eR\to\eR$ {\em is associated with the set A} (or {\em corresponds to $A$}) if it is nonnegative and $A$ has the following tube formula:
\begin{equation}
|A_t|=t^{N-D}(G(\log(1/t))+o(1))\textrm{ as }t\to0^+,
\end{equation}
with $0<\liminf_{\tau\to+\ty} G(\tau)\le\limsup_{\tau\to+\ty} G(\tau)<\ty$.

\medskip

In addition, we say that $A$ is a {\em transcendentally $n$-quasiperiodic set} if the corresponding function $G=G(\tau)$ is transcendentally 
quasiperiodic. We say that {\em $A$ is algebraically $n$-quasiperiodic} if the corresponding function $G$ is algebraically $n$-quasiperiodic. The smallest possible value of $n$ is called the {\em order of quasiperiodicity of $A$} (and {\em of $G$}).
\end{defn}

\medskip

The following result, which has a variety of generalizations, as will be briefly explained below, provides a construction of transcendentally $2$-quasi\-peri\-odic fractal sets. Its proof is based on the classical Gel'fond--Schneider theorem (as described in~\cite{gelfond}) from transcendental number theory.

\medskip

\begin{theorem}[\cite{fzf}]\label{trans}
Let $A_1=C^{(m_1,a_1)}\st I_1$ and $A_2=C^{(m_2,a_2)}\st I_2$ be two generalized Cantor 
sets, contained in two unit closed intervals $I_1$ and $I_2$ with disjoint interiors and such that their box dimensions coincide. 
Let $\{p_1,p_2,\dots,p_k\}$ be the set of all distinct prime factors of $m_1$ and $m_2$, and write
\begin{equation}
m_1=p_1^{\alpha_1}p_2^{\alpha_2}\dots p_k^{\alpha_k},\quad m_2=p_1^{\beta_1}p_2^{\beta_2}\dots p_k^{\beta_k},
\end{equation}
where $\alpha_i,\beta_i\in\eN\cup\{0\}$\label{n_0} for $i=1,\ldots,k$. If the exponent vectors
\begin{equation}
(\alpha_1,\alpha_2,\dots,\alpha_k)\,\,\qs\mathrm{and}\qs\,\,(\beta_1,\beta_2,\dots,\beta_k),
\end{equation}
corresponding to $m_1$ and $m_2$,
are linearly independent over the rationals, then the compact set $A:=A_1\cup A_2$
is transcendentally $2$-quasiperiodic.

\medskip

Moreover, the distance zeta function can be meromorphically extended to the whole complex plane, and we have that
$D(\zeta_{A})=D$. The set $\dim_{PC}A$ of principal complex dimensions of $A$ is given by
\begin{equation}\label{511/2}
\dim_{PC} A=
D+\Big(\frac{2\pi}{T_1}\Ze\cup \frac{2\pi}{T_2}\Ze \Big)\I.
\end{equation}
Besides $(\dim_{PC}A)\cup\{0\}$, there are no other poles of the distance zeta function~$\zeta_{A}$.
\end{theorem}

\medskip

This result can be considerably extended by using Baker's theorem \cite[Theorem~2.1]{baker} from transcendental number theory.
In short, we construct transcendentally $n$-quasiperiodic fractal sets for any given integer $n\ge2$ (and even for $n=\ty$, in the sense of Definition~\ref{quasipty} below).
The corresponding extension is provided in Chapter~3 of~\cite{fzf}; see also Theorem~\ref{ty} below and the discussion preceding it.

\medskip

\begin{theorem}[\cite{fzf}]\label{quasi1}
Let $n\in\eN$ with $n\geq 2$.
Assume that $A_j=C^{(m_j,a_j)}\st I_j$, $j=1,\dots,n$, are generalized Cantor sets $($in the sense of Definition~\ref{Cma}$)$ such that their box dimensions are equal to a fixed number 
$D\in(0,1)$. Also assume that they are contained in closed unit intervals~$I_j$ with pairwise disjoint interiors.
Let $T_i:=\log(1/a_i)$ be the associated periods, and $G_i$ be the corresponding $T_i$-periodic functions, for $i=1,\dots, n$.
Let $\{p_j:j=1,\dots,k\}$ be the union of all distinct prime factors  which appear in the prime factorization of each of the integers $m_i$, for $i=1,\dots, n$; that is, $m_i=p_1^{\a_{i1}}\dots p_k^{\a_{ik}}$, where $\a_{ij}\in\eN\cup\{0\}$.
If the exponent vectors of the numbers $m_i$,
\begin{equation}\label{2.1.111/2}
e_i=(\a_{i1},\dots,\a_{ik}),\quad i=1,\dots,n,
\end{equation}
are linearly independent over the rationals, then the compact set
$$A:=A_1\cup\dots\cup A_n\subseteq\eR$$
is transcendentally $n$-quasiperiodic.

\medskip

Moreover, the distance zeta function $\zeta_A$ can be meromorphically extended to the whole complex plane, and $D(\zeta_A)=D$.
The set $\dim_{PC}A$ of principal complex dimensions of $A$ is given by
\begin{equation}\label{521/2}
\dim_{PC} A=D+\Big(\bigcup_{i=1}^n\frac{2\pi}{T_i}\Ze\Big)\I.
\end{equation}
Besides $(\dim_{PC}A)\cup\{0\}$, there are no other poles of the distance zeta function~$\zeta_A$.
\end{theorem}

\medskip

In Section~\ref{relative} (Theorem~\ref{ty}), we will even construct a set which is transcendentally $\infty$-quasiperiodic (and hyperfractal).

\section{Zeta functions of relative fractal drums}\label{relative}

The notion of a relative fractal drum, studied in detail in Chapter~4 of~\cite{fzf}, is a convenient and flexible tool which enables us to naturally extend already existing notions pertaining to bounded fractal strings and bounded fractal drums.
It is noteworthy that the box dimension of a given relative fractal drum may be negative.

\medskip

We begin by giving the analog (in this more general context) of Definition~\ref{defn} for the distance zeta function $\zeta_A$ of a bounded set $A\subset\eR^N$.
(The corresponding analog of Definition~\ref{zeta_tilde} for the tube zeta function $\widetilde{\zeta}_A$ of a bounded set is given in part $(b)$ of Remark~\ref{5.11/2} below.)

\medskip

\begin{defn}[\cite{fzf}]\label{zeta_r}\label{drum} Let $\Omega$ be an open subset of $\eR^N$, not necessarily bounded, but of finite $N$-dimensional Lebesgue measure (or ``volume'').
Furthermore, let $A\subseteq\eR^N$, also possibly unbounded, such that $\Omega$ is contained in $A_\delta$ for some $\delta>0$.
The {\em distance zeta function $\zeta_A(\,\cdot\,,\Omega)$ of $A$ relative to $\Omega$} (or the {\em relative distance 
zeta function})
 is defined by
\begin{equation}\label{rel_dist_zeta}
\zeta_A(s,\Omega)=\int_{\Omega} d(x,A)^{s-N}\D x,
\end{equation}
for all $s\in\Ce$ with $\re s$ sufficiently large. We  call the ordered pair $(A,\Omega)$, appearing in Definition~\ref{zeta_r},  
a {\em relative fractal drum} (RFD). 
Therefore, we shall also use the phrase {\em zeta functions of relative fractal drums} instead of relative zeta functions.
\end{defn}

\medskip

\begin{remark}
We point out that if we replace the domain of integration $\O$ in~\eqref{rel_dist_zeta} with $A_\d\cap\O$ for $\d>0$, then similarly as before, the dependence of the relative distance zeta function on the number $\delta$ is inessential.
The condition that $\O\subseteq A_\d$ for some $\d>0$ is of technical nature and ensures that the function $x\mapsto d(x,A)$ is bounded for $x\in\O$.
If $\O$ does not satisfy this condition we can still replace it with $A_\d\cap\O$ for some fixed $\d>0$ and apply the theory.\footnote{Since then, $\O\setminus A_\d$ and $A$ are a positive distance apart, this replacement will not affect the relative box dimension of $(A,\O)$ introduced just below.}

\end{remark}

\medskip

\begin{remark}\label{5.11/2}
$(a)$ More generally, one does not need to assume that $\O$ has finite volume.
Instead, it suffices to assume that for some $\delta>0$, $A_{\d}\cap\O$ has finite volume, and correspondingly, to define
\begin{equation}\label{rel_dist_zeta_d}
\zeta_A(s,\Omega)=\int_{A_\d\cap\Omega} d(x,A)^{s-N}\D x,
\end{equation}
for all $s\in\Ce$ such that $\re s$ is sufficiently large.
Furthermore, it is not really necessary to assume that $\O$ is an open subset of $\eR^N$ (one may simply assume that $\O\subseteq\eR^N$ is Lebesgue measurable), although the latter assumption justifies the term ``relative fractal drum''.
We will use this refinement of Definition~\ref{zeta_r} when discussing the new notion of {\em $($pointwise$)$ local} fractal zeta function.

\medskip

$(b)$ An entirely analogous comment can be made about the {\em relative tube zeta function} of $(A,\O)$, defined (under the just mentioned hypotheses in part~$(a)$ of this remark) by
\begin{equation}\label{401/2}
\widetilde{\zeta}_A(s,\O)=\int_0^{\delta}t^{s-N-1}|A_t\cap\O|\,\D t,
\end{equation}
for all $s\in\Ce$ with $\re s$ sufficiently large.
Finally, we note that the exact counterpart of Remark~\ref{2.71/2} above holds for the relative (as opposed to ``absolute'') tube and distance zeta functions $\widetilde{\zeta}_A(\,\cdot\,,\O)$ and $\zeta_A(\,\cdot\,,\O)$, as is stated in part $(c)$ of the present remark.

\medskip

$(c)$ In light of the counterpart of Remark~\ref{2.71/2} in the present context, we always have that
\begin{equation}\label{403/4}
D(\zeta_A(\,\cdot\,,\O))=D(\widetilde{\zeta}_A(\,\cdot\,,\O))
\end{equation}
and, provided $D(\zeta_A(\,\cdot\,,\O))<N$, we also have that
\begin{equation}\label{404/5}
D_{hol}(\zeta_A(\,\cdot\,,\O))=D_{hol}(\widetilde{\zeta}_A(\,\cdot\,,\O)).
\end{equation}
In other words, the relative distance and tube zeta functions have the same abscissa of (absolute) convergence and, if $D(\zeta_A(\,\cdot\,,\O))<N$, also the same abscissa of holomorphic convergence (see the discussion preceding Theorem~\ref{an} above); we refer to Theorem~\ref{an_rel} along with Corollaries~\ref{an1r} and~\ref{2.31/2r} below for further information about these abscissae of convergence and their relationship.

\medskip

$(d)$ In order to recover the usual definition of the (absolute) distance and tube zeta functions (as in Definition~\ref{defn} or Definition~\ref{zeta_tilde}, respectively), it suffices to set $\O:=A_{\d}$ in formula~\eqref{rel_dist_zeta} or~\eqref{401/2}, respectively.
\end{remark}

\medskip

Next, we provide some additional relative analogs of well-known definitions (recalled in Section~\ref{intro} above).

\medskip

First, for any real number $r$, we define the {\em upper $r$-dimensional Minkowski content of $A$ relative to $\Omega$}
(or {\em the upper relative Minkowski content}, or {\em the upper Minkowski content of the relative fractal drum $(A,\Omega)$}) by
\begin{equation}\label{minkrel}
\mathcal{M}^{*r}(A,\Omega)=\limsup_{t\to0^+}\frac{|A_t\cap\Omega|}{t^{N-r}}, 
\end{equation}
and then we proceed in the usual way:
\begin{equation}\label{dimrel}
\begin{aligned}
\ov\dim_B(A,\Omega)&=\inf\{r\in\eR:\mathcal{M}^{*r}(A,\Omega)=0\} \\
&=\sup\{r\in\eR:\mathcal{M}^{*r}(A,\Omega)=+\ty\}.
\end{aligned}
\end{equation}
We call it the {\em relative upper box dimension}
 $($\rm{or} {\em relative Minkowski dimension}$)$ of $A$ with respect to $\Omega$ (or else the {\em relative upper box dimension of $(A,\Omega)$}).
Note that $\ov\dim_B(A,\Omega)\in[-\ty,N]$, and the values can indeed be negative, even equal to $-\ty$; 
see Proposition~\ref{ndim} and Corollary~\ref{flat}.
Also note that for these definitions to make sense, it is not necessary that $\O\subseteq A_\d$ for some $\d>0.$

\medskip

The value $\mathcal{M}_*^{r}(A,\Omega)$ of the {\em lower $r$-dimensional Minkowski content} of $(A,\Omega)$, is defined as in \eqref{minkrel}, except for a lower instead of an upper limit.
Analogously as in \eqref{dimrel}, we define the {\em relative lower box $(${\rm or} Minkowski$)$ dimension} of $(A,\Omega)$:
\begin{equation}\label{dimrel2}
\begin{aligned}
\underline\dim_B(A,\Omega)&=\inf\{r\in\eR:\mathcal{M}_*^r(A,\Omega)=0\}\\
&=\sup\{r\in\eR:\mathcal{M}_*^r(A,\Omega)=+\infty\}.
\end{aligned}
\end{equation} 
Furthermore, when $\underline\dim_B(A,\Omega)=\ov\dim_B(A,\Omega)$, we denote by
$
\dim_B(A,\Omega)
$ 
this common value and then say that the {\em relative box $(${\rm or} Minkowski$)$ dimension\label{rel_box_dim} $\dim_B(A,\Omega)$ exists}.
If $0<\mathcal{M}_*^D(A,\Omega)\le\mathcal{M}^{*D}(A,\Omega)<\ty$, we say that the relative 
fractal drum $(A,\O)$ is {\em Minkowski nondegenerate}.\label{nondeg_rel}
It then follows that $\dim_B(A,\O)$ exists and is equal to $D$.

\medskip

Moreover, if $\mathcal{M}_*^D(A,\Omega)=\mathcal{M}^{*D}(A,\Omega)$, this common value is denoted by $\mathcal{M}^D(A,\Omega)$ and called the {\em relative 
Minkowski content} of $(A,\O)$.
Finally, if $\mathcal{M}^D(A,\Omega)$ exists and is different from $0$ and $\ty$ (in which case $\dim_B(A,\Omega)$ exists and then necessarily $D=\dim_B(A,\Omega)$), we say that the relative fractal drum $(A,\Omega)$ is {\em Minkowski 
measurable}.
Various examples and properties of relative box dimensions can be found in [Lap1--3], 
[LapPo1--3], 
\cite{lapidushe}, [Lap--vFr1--3], 
\cite{rae}, [LaPe2--3], 
[LapPeWi1--2] 
and [LapRa\v Zu1--7].

\medskip

We now state the first main result for relative distance zeta functions, which extends Theorem \ref{an} to relative fractal drums.
(Naturally, this result has an exact analog for relative tube functions, without the need for the condition $D<N$ from part~$(c)$.)

\medskip

\begin{theorem}[\cite{fzf}]\label{an_rel}
Let $\Omega$ be a $($nonempty$)$ open subset of $\eR^N$ of finite $N$-dimensional Lebesgue measure, and let $A\st\eR^N$.\footnote{More generally, let $\O$ be any (nonempty) Lebesgue measurable subset of $\eR^N$ such that $|A_{\d_1}\cap\O|<\ty$, for some $\d_1>0$.
In that case, we have to use the alternative definition of $\zeta_A(\,\cdot\,,\O)$ given by~\eqref{rel_dist_zeta_d} in part $(a)$ of Remark \ref{5.11/2}.} Then the following properties hold$:$

\bigskip 

$(a)$ The relative distance zeta function $\zeta_A(s,\Omega)$ is holomorphic in the half-plane $\{\re s>\overline{\dim}_B(A,\Omega)\}$.

\bigskip

$(b)$ In part $(a)$, the half-plane $\{\re s>\ov\dim_B(A,\O)\}$ is optimal, from the point of view of the convergence $($and therefore, also of the absolute convergence$)$ of the Lebesgue integral defining $\zeta_A(s,\Omega)$ in~\eqref{rel_dist_zeta}$;$ i.e., the equality~\eqref{ddimr} in Corollary~\ref{an1r} below holds.
Furthermore, $\zeta_A(s,\O)$ is still given by~\eqref{rel_dist_zeta} for $\re s>\ov{\dim}_B(A,\O);$ i.e., the second inequality holds in~\eqref{51/4r} of Corollary~\ref{an1r} below.


\bigskip

$(c)$ If the relative box $($or Minkowski$)$ dimension $D:=\dim_B(A,\O)$ exists, $D<N$, and $\M_*^D(A,\O)>0$, then $\zeta_A(s,\O)\to+\ty$ as $s\in\eR$
converges to $D$ from the right.
Therefore, the equalities~\eqref{51/2r} in Corollary~\ref{2.31/2r} below hold.   
\end{theorem}

\medskip

Much as before in Section~\ref{disttube}, we can state the next two corollaries which tell us, in particular, how the relative upper box dimension of $(A,\O)$ can be deduced from the relative distance zeta function of $(A,\O)$.

\medskip

\begin{cor}[\cite{fzf}]\label{an1r}
If $(A,\O)$ is a relative fractal drum in $\eR^N$, then
\begin{equation}\label{ddimr}
\ov\dim_B(A,\O)=D(\zeta_A(\,\cdot\,,\O)).
\end{equation}
Furthermore, in general, we have
\begin{equation}\label{51/4r}
-\ty\leq D_{hol}(\zeta_A(\,\cdot\,,\O))\leq D(\zeta_A(\,\cdot\,,\O))=\ov{\dim}_B(A,\O).
\end{equation}
\end{cor}

\medskip

\begin{cor}[\cite{fzf}]\label{2.31/2r}
Let $(A,\O)$ be a relative fractal drum in $\eR^N$ which satisfies the hypotheses of part $(c)$ of Theorem~\ref{an_rel}.
Then, we have
\begin{equation}\label{51/2r}
\ov{\dim}_B(A,\O)=D(\zeta_A(\,\cdot\,,\O))=D_{hol}(\zeta_A(\,\cdot\,,\O)).
\end{equation}
\end{cor}

\medskip

\begin{remark}\label{5.41/2}
Recall from Remark~\ref{2.41/2} that we do not know whether there exist natural (and nontrivial) fractal bounded sets (and, therefore, relative fractal drums) for which the identity~\eqref{51/2r} in Corollary~\ref{2.31/2r} does not hold; that is, in light of Corollary~\ref{an1r}, for which the second inequality in~\eqref{51/4r} of Corollary~\ref{an1r} is strict:
\begin{equation}\label{461/4}
D_{hol}(\zeta_A(\,\cdot\,,\O))< D(\zeta_A(\,\cdot\,,\O))=\ov{\dim}_B(A,\O).
\end{equation}
Since relative fractal drums can have negative dimensions, it would be interesting to find a natural class of relative fractal drums for which the strict inequality in~\eqref{461/4} still holds but $\ov{\dim}_B(A,\O)=D(\zeta_A(\,\cdot\,,\O))$ (and hence also $D_{hol}(\zeta_A(\,\cdot\,,\O))$, in light of inequality~\eqref{51/4r}) is strictly negative.
Of course, provided $D:=\dim_B(A,\O)$ exists and $D<N$, in light of Corollary~\ref{2.31/2r} and part $(c)$ of Theorem~\ref{an_rel}, they should also have the property that $\mathcal{M}_*^D(A,\O)=0$.
\end{remark}

\medskip

The following proposition provides a class of relative fractal drums in the plane, the box dimension of which is strictly negative.

\medskip

\begin{prop}[\cite{fzf}]\label{ndim}
Let $A=\{(0,0)\}$ and
$$
\Omega=\{(x,y)\in\eR^2:0<y<x^\alpha,\,\,x\in(0,1)\}, 
$$
where $\alpha>1$.
Then the relative fractal drum $(A,\O)$ has a negative box dimension. More specifically, $\dim_BA$ exists, the relative fractal drum $(A,\O)$ is Minkowski measurable and
\begin{equation}
\begin{gathered}
\dim_B(A,\Omega)=D(\zeta_A(\,\cdot\,,\Omega))=1-\alpha<0,\\
\mathcal{M}^{1-\alpha}(A,\Omega)=\frac1{1+\alpha},\\
\end{gathered}
\end{equation}
and $\zeta_A(\,\cdot\,,\Omega)$ can be meromorphically extended at least to $\{\re s>3(1-\a)\}$.
Furthermore, $D=1-\alpha$ is a simple pole of $\zeta_A(\,\cdot\,,\Omega)$.
\end{prop}

\medskip

Furthermore, it is even possible to construct nontrivial relative fractal drums for which the corresponding box dimension is equal to $-\ty$.

\medskip

\begin{cor}[\cite{fzf}]\label{flat}
Let $A=\{(0,0)\}$ and 
$$
\Omega'=\{(x,y)\in\eR^2:0<y<{\E}^{-1/x},\,\,0<x<1\}.
$$
Then
\begin{equation}
\dim_B(A,\Omega')=D(\zeta_A(\,\cdot\,,\Omega')=-\ty.
\end{equation}
\end{cor}

\medskip

The relative distance zeta function has a nice scaling property, which can be useful in the study of relative fractal sprays in Euclidean spaces; see \cite{fzf}.

\medskip

\begin{theorem}[Scaling property of relative zeta functions; \cite{fzf}]\label{scaling}
Let $\zeta_A(\,\cdot\,,\Omega)$ be the relative distance zeta function. Then, for any positive real number $\lambda$, we have $D(\zeta_{\lambda A}(\,\cdot\,,\lambda\O))=D(\zeta_A(\,\cdot\,,\O))=\overline{\dim}_B(A,\O)$ and
\begin{equation}\label{zeta_scaled}
\zeta_{\lambda A}(s,\lambda\Omega)=\lambda^s\zeta_A(s,\Omega),
\end{equation}
for $\re s>\overline{\dim}_B(A,\O)$. 
\end{theorem}

\medskip

\begin{example}\label{sierpinski_carpetr} (Relative Sierpi\'nski carpet)
Let $A$ be the Sierpi\'nski carpet based on the unit square $\O$. Let $(A,\O)$ be the corresponding {\em relative Serpi\'nski carpet}, with $\Omega$ being the unit square.
In this case, the relative distance zeta function of $(A,\O)$ has a meromorphic continuation to the entire complex plane given by
\begin{equation}\label{10.111/4}
\zeta_A(s,\Omega)=\frac{8}{2^ss(s-1)(3^s-8)},\q\mbox{\rm for all}\q s\in\Ce.
\end{equation}

\medskip

Here, the relative box dimension of $(A,\O)$ coincides with its usual box dimension, namely, $\log_38$ (which also coincides with the Minkowski and Hausdorff dimensions of the standard Sierpi\'nski carpet $A$). Moreover, the set $\dim_{PC}(A,\O)$ of relative principal complex dimensions of the Sierpi\'nski carpet $(A,\O)$ is given by 
\begin{equation}\label{501/4}
\dim_{PC}(A,\O)=\log_38+\mathbf{p}{\I}\Ze,
\end{equation}
where 
$\mathbf{p}:=2\pi/\log 3$ 
is the oscillatory period of the Sierpi\'nski carpet $A$.

\medskip

Observe that it follows immediately from $(\ref{10.111/4})$ that the set $\po(\zeta_A(\,\cdot\,,\Omega))$ of all relative complex dimensions of the Sierpi\'nski carpet $A$ (with respect to the unit square $\Omega$) is given by 
\begin{equation}\label{501/2}
\po(\zeta_A(\,\cdot\,,\Omega))=\dim_{PC}A\cup\{0,1\}=(\log_38+\mathbf{p}{\I}\Ze)\cup\{0,1\},
\end{equation}
where $\{0,1\}$ can be viewed as the set of `integer dimensions' of $A$ (in the sense of 
[{LapPe2--3}] 
and 
[LaPeWi1--2], see also \cite[Section~13.1]{lapidusfrank12}).\footnote{Furthermore, each of these relative complex dimensions is simple (i.e., is a simple pole of $\zeta_A(\,\cdot\,,\O)$).} Interestingly, these are exactly the complex dimensions which one would expect to be associated with $A$, according to the theory developed in 
[{LapPe2--3}] 
 and
[{LapPeWi1--2}] (as described in \cite[Section 13.1]{lapidusfrank12})
via self-similar tilings (or sprays) and associated tubular zeta functions. 

\medskip

In light of \eqref{10.111/4} and~\eqref{501/4}, the residue of the distance zeta function of the relative Sierpi\'nski carpet $(A,\O)$ computed at any principal pole $s_k:=\log_38+\mathbf{p}{\I}k$, $k\in\Ze$, is given by
$$
\res(\zeta_A(\,\cdot\,,\O),s_k)=\frac{2^{-s_k}}{(\log3)s_k(s_k-1)}.
$$
In particular,
$$
|\res(\zeta_A(\,\cdot\,,\O),s_k)|\sim \frac{2^{-D}}{D\log3}\,k^{-2}\q\mbox{as\q$k\to\pm\ty$,}
$$
where $D:=\log_38$.
It follows, in particular, that
$$
\res(\zeta_A(\,\cdot\,,\O),s_k)\to 0\q\textrm{as}\q k\to\pm\ty.
$$
\end{example}\label{sierp_end}

\medskip

The notion of $n$-quasiperiodicity of relative fractal drums, introduced (for bounded sets) in Definition \ref{quasiperiodic} above, can be extended to the case where $n=\ty$. To this end, we first define  $\ty$-quasiperiodic functions, by modifying Definition \ref{quasip}. We shall need the space $\eR^\ty_b$ of bounded sequences of real numbers.
We note that the material in the rest of this section is excerpted from~\cite[Section~4.6]{fzf}.

\medskip

\begin{defn}[\cite{fzf}]\label{quasipty}
We say that a function $G=G(\tau):\eR\to\eR$ is {\em transcen\-dentally 
$\ty$-quasiperiodic}
 if it is of the form $G(\tau)=H(\tau,\tau,\dots)$,
where $H:\eR^\ty_b\to\eR$  is a function which is nonconstant and $T_k$-periodic in its $k$-th component, for each $k\in\eN$, and the set of periods $\{T_k:k\in\eN\}$ is algebraically (and hence, rationally) independent.\footnote{An infinite subset $B$ of a vector space $X$ is said to be linearly independent (over a given field $\mathbb{K}$) if any finite subset of $B$ is linearly independent (over $\mathbb{K}$).
Here, according to the context, $\mathbb{K}$ stands for either the field of rational numbers or the field of algebraic numbers.
Recall that the {\em algebraic numbers} are the roots of polynomial equations with integer (or, equivalently, rational) coefficients.
So that the field of algebraic numbers (considered as a subfield of $\Ce$) can be viewed as a realization of the algebraic closure of $\mathbb{Q}$, the field of rational numbers.
By definition, the set of {\em transcendental numbers} (\cite{baker}) is the complement in $\eR$ of the set of algebraic numbers.}
The values of $T_k$ are called the 
{\em quasiperiods of $G$}. If the set $\{T_k:k\in\eN\}$ of quasiperiods is rationally independent and algebraically dependent (i.e., there exists a finite subset of periods which is algebraically dependent), we say that {\em $G$ is algebraically $\ty$-quasiperiodic}. 
\end{defn}

\medskip

Now, we can introduce the definition of an $n$-quasiperiodic relative fractal drum analogously to Definition \ref{quasiperiodic}, for $n\ge2$ or $n=\ty$.

\medskip

\begin{defn}[\cite{fzf}]
A relative fractal drum $(A,\O)$ is said to be {\em transcendentally $n$-quasiperiodic}, where $n\ge2$ or $n=\ty$, if the corresponding relative tube formula has the form
\begin{equation}
|A_t\cap\O|=t^{N-D}(G(\log(1/t))+o(1))\textrm{ as }t\to0^+,
\end{equation}
where the function $G:\eR\to\eR$ is nonnegative, $0<\liminf_{\tau\to+\ty} G(\tau)\le\limsup_{\tau\to+\ty} G(\tau)<\ty$, and it is transcendentally $n$-quasiperiodic in the sense of Definition \ref{quasip} for $n\ge2$ and in the sense of Definition \ref{quasipty} for $n=\ty$. We can analogously define {\em algebraically $n$-quasiperiodic relative fractal drums}, where $n\ge2$ or $n=\ty$. The smallest possible value of $n$ is called the {\em order of quasiperiodicity of the relative fractal drum}.
\end{defn}

\medskip

If $A$ is a given bounded subset of $\eR^N$, then it can be identified with a relative fractal drum $(A,A_\d)$, where $\d$ is an arbitrary fixed positive real number.
That is, we let $\O:=A_{\d}$ in Definition~\ref{zeta_r} of a relative fractal drum; see also part $(d)$ of Remark~\ref{5.11/2} above.
We then say that the set {\em $A$ is transcendentally algebraically $\ty$-quasiperiodic} if the relative fractal drum $(A,A_\d)$ is $\ty$-quasiperiodic.
Recall that we have already introduced the definition of $n$-quasiperiodic sets for any integer $n\ge2$; see Definition \ref{quasiperiodic}.

\medskip

The notion of $n$-quasiperiodicity, where $n\ge2$ or $n=\ty$, can also be introduced for arbitrary bounded fractal strings and bounded subsets of Euclidean spaces.

\medskip

More information and results about quasiperiodic sets and drums can be found in~\cite{fzf} and [LapRa\v Zu2--4].

\medskip

\begin{defn}[\cite{fzf}]\label{tytr_quasiperiodic}
Any bounded fractal string $\mathcal{L}=(l_j)_{j\ge1}$ can be naturally identified with a relative fractal drum $(A_{\mathcal{L}},\Omega_{\mathcal{L}})$ in $\eR$,
where 
$$
A_{\mathcal{L}}:=\bigg\{a_k:=\sum_{j\ge k}l_j:k\ge1\bigg\},\q \O_{\mathcal{L}}:=\bigcup_{k=1}^\ty(a_{k+1},a_k),
$$ 
with $|\O_{\mathcal{L}}|=\sum_{j=1}^\ty l_j<\ty$.
We say that {\em a bounded fractal string $\mathcal{L}=(l_j)_{j\ge1}$ is $($transcendentally$)$ algebraically
$n$-quasiperiodic}, where $n\ge2$ or $n=\ty$,
if the corresponding relative fractal drum $(A_{\mathcal{L}},\O_{\mathcal{L}})$ is (transcendentally) algebraically $n$-quasiperiodic.
\end{defn}

\medskip

\begin{remark}\label{footnote25}
For the purpose of this paper, and following~[Lap--vFr1--3], 
 a {\em screen} $S$ is defined as the graph of a bounded, Lipschitz real-valued function $t\mapsto S(t)$ $(t\in\eR)$, with the horizontal and vertical axes interchanged: $S=\{S(t)+\I t\,:\,t\in\eR\}\subseteq\Ce$, where $\I:=\sqrt{-1}$.
We assume that $S$ is located to the left of the critical line $\{\re s=D\}$.
For example, $S$ could be a vertical line $\{\re s=\alpha\}$, with $-\ty<\alpha\leq D$.
Moreover, the {\em window} $W$ associated with the screen $S$ is the closed subset of $\Ce$ defined by $W:=\{s\in\Ce\,:\,\re s\geq S(\im s)\}$.
(See~\cite[Section~5.1]{lapidusfrank12} for more details.)
\end{remark}

\medskip

\begin{defn}[\cite{fzf}]\label{analytically}\label{hyperfractal}
Let $A$ be a bounded subset of $\eR^N$ and let $D:=D(\zeta_A)=\ov{\dim}_BA$.

\medskip

$(i)$ The set $A$ is a {\em hyperfractal} (or is {\em hyperfractal}) if there is a screen $S$ along which the associated tube (or equivalently, if $D<N$, distance) zeta function has a 
natural boundary. This means that the zeta function cannot be meromorphically extended to an open neighborhood of the associated window $W$ (with associated screen denoted by $S$), in the terminology of Remark~\ref{footnote25}.\footnote{This definition is closely related to the notion of fractality given in \cite[Sections 12.1.1 and 12.1.2, including Figures 12.1--12.3]{lapidusfrank12}; see Remark~\ref{5.151/2}.}

\medskip

$(ii)$ The set $A$ is a {\em strong hyperfractal} (or is {\em strongly hyperfractal}) if the 
critical line $\{\re s =D\}$ is a 
natural boundary
of the associated zeta function; that is, if we can choose $S = \{\re s = D\}$ in~$(i)$.

\medskip

$(iii)$ Finally, the set $A$ is {\em maximally hyperfractal} if it is strongly  hyperfractal and every point of the critical line is a nonremovable singularity of (the meromorphic continuation of) the zeta function.\footnote{Recall that the fractal zeta function $\zeta_A$ is holomorphic on $\{\re s>\ov{\dim}_BA\}$.}

\medskip

An analogous definition can be provided (in the obvious manner) where instead of $A$, we have a fractal string $\mathcal{L}=(l_j)_{j\ge1}$ in $\eR$ or a relative fractal drum $(A,\O)$ in~$\eR^N$.
\end{defn}

\medskip

\begin{remark}\label{5.151/2}
In [Lap--vFr1--3], 
a geometric object is said to be ``fractal'' if the associated zeta function has at least one nonreal complex dimension (with positive real part).
(See \cite[Sections~12.1 and~12.2]{lapidusfrank12}) for a detailed discussion.
In \cite{lapidusfrank06,lapidusfrank12}, in order, in particular, to take into account some possible situations pertaining to random fractals (\cite{HamLa}), the definition of fractality (within the context of the theory of complex dimensions) was extended so as to allow for the case described in part $(i)$ of Definition~\ref{hyperfractal} just above, namely, the existence of a natural boundary along a screen.
(See \cite[Subsection~13.4.3]{lapidusfrank12}.)

We note that in \cite{lapidusfrank12} (and the other aforementioned references), the term ``hyperfractal'' was not used to refer to case $(i)$.
More important, except for fractal strings and in very special higher-dimensional situations (such as suitable fractal sprays), one did not have (as we now do) a general definition of ``fractal zeta function'' associated with an arbitrary bounded subset of $\eR^N$, with $N\geq 1$.
Therefore, we can now define the ``fractality'' of any bounded subset of $\eR^N$ (including Julia sets and the Mandelbrot set) and, more generally, of any relative fractal drum, by the presence of a nonreal complex dimension (with a positive real part) or by the ``hyperfractality'' (in the sense of part $(i)$ of Definition~\ref{hyperfractal}) of the geometric object.
Here, ``complex dimension'' is understood as a (visible) pole of the associated fractal zeta function (the distance or tube zeta function of a bounded subset or a relative fractal drum of $\eR^N$, or else, as was the case in most of \cite{lapidusfrank12}, the geometric zeta function of a fractal string).  
\end{remark}

\medskip

The following theorem (from~\cite[Section~4.6]{fzf}) shows that there exists an effectively constructible maximally hyperfractal set.
This result can easily be extended to any dimension $N\geq 1$.
(See Corollary~\ref{5.171/2} below.)
Moreover, in light of Definition~\ref{hyperfractal} and Remark~\ref{5.151/2} above, it provides a class of (deterministic) examples that are as ``fractal'' as possible. 

\medskip

\begin{theorem}[\cite{fzf}]\label{ty}
There exists a bounded fractal string $\mathcal{L}$ which is maximally hyperfractal and $\ty$-quasiperiodic.
In particular, the corresponding bounded set $A_{\mathcal{L}}$ is also maximally hyperfractal and transcendentally $\ty$-quasiperiodic.
\end{theorem}

\medskip

The maximally hyperfractal bounded string $\mathcal{L}$ in Theorem \ref{ty} is obtained as the disjoint union of a sequence of (suitably scaled) two-parameter generalized Cantor sets $C^{(m_j,a_j)}$, $j\in\eN$ (viewed as fractal strings), with carefully chosen parameters $m_j$ and $a_j$,
such that the associated set $\{T_j=\log m_j:j\in\eN\}$ of quasiperiods is algebraically (and hence, rationally) linearly independent.

\medskip

\begin{cor}[\cite{fzf}]\label{5.171/2}
The counterpart of Theorem~\ref{ty} holds in any dimension $N\geq 1$, both for relative fractal drums of $\eR^N$ and for bounded subsets of $\eR^N$.
\end{cor}

\medskip

We close this section by discussing several variants of a new notion of pointwise (or local) fractal zeta function.
However, we should point out that the new notions discussed in Definition~\ref{5.161/4}, Remark~\ref{5.161/2} and Remark~\ref{5.164/5}, as well as in Exercise \ref{5.163/4} below, are still evolving and may not yet be presented in their definitive form.
Nevertheless, even at this preliminary stage, they seem to be of significant interest in many situations and to lead to the correct (or expected) results.

\medskip

\begin{defn}\label{5.161/4}
Consider an arbitrary (nonempty) Borel subset $\O$ of $\eR^N$ (which is neither assumed to have finite volume nor to be open, in agreement with part $(a)$ of Remark~\ref{5.11/2}).
Let us fix $x\in\ov\O$. We are interested in the fractal properties of $\O$ near $x$. To this end,  let $A:=B_r(x)$, for some $r>0$, where $A$ is the ball of center $x$ and radius $r$.
Then, the {\em $($pointwise$)$ local distance zeta function of $\O$ at $x$}, denoted by $\zeta_x(\,\cdot\,,\O)$, is defined as the distance zeta function $\zeta_A(\,\cdot\,,\O)$ of the relative fractal drum $(A,\O):=(B_r(x),\O)$, in the sense of Remark~\ref{5.11/2} (note that $A_\d=B_{r+\d}(x)$):
\begin{equation}\label{681/4}
\begin{aligned}
\zeta_x(s,\O)&:=\zeta_A(s,\O)=\zeta_{B_r(x)}(s,\O)\\
&\phantom{:}=\int_{B_{r+\d}(x)\cap\O}d(y,B_r(x))^{s-N} \D y\\
&=\int_{B_{r,r+\d}(x)\cap\O}d(y,B_r(x))^{s-N},
\end{aligned}
\end{equation}
for all $s\in\Ce$ such that $\re s>N$,
where $\d>0$ is fixed in advance, and $B_{r,r+\d}(x):=B_{r+\d}(x)\setminus B_r(x)$ is the annulus centered at $x$ with inner radius $r$ and outer $r+\d$.\footnote{The last equality in Equation \eqref{681/4} is due to the fact that $d(y,B_r(x))=0$ for all $y\in B_r(x)$.}
Note that if $|\O|=0$, then this definition is of no use, since in that case, we have that $\zeta_x(s,\O)\equiv0$.

\medskip

We define similarly $\widetilde{\zeta}_x(\,\cdot\,,\O)$, the {\em $($pointwise$)$ local tube zeta function of} $\O$, as the tube zeta function of the same relative fractal drum $(A,\O):=(B_r(x),\O)$:
\begin{equation}\label{682/3}
\begin{aligned}
\widetilde{\zeta}_x(s,\O)&:=\widetilde{\zeta}_A(s,A_\d\cap\O)=\widetilde{\zeta}_{B_r(x)}(s,\O)\\
&\phantom{:}=\int_0^\d t^{N-s-1}|B_{r+t}(x)\cap\O|\,\D t.
\end{aligned}
\end{equation}
\end{defn}
Again, in the case when $|\O|=0$, we have $\widetilde{\zeta}_x(s,\O)\equiv0$. Therefore, the definition of the local tube zeta function can be useful only when $|\O|>0$.

\medskip

\begin{remark}\label{5.161/2}
$(a)$~In the above definition, we should really denote the local fractal zeta functions ${\zeta}_x(\,\cdot\,,\O)$ and $\widetilde{\zeta}_x(\,\cdot\,,\O)$ by ${\zeta}_{x,r}(\,\cdot\,,\O)$ and $\widetilde{\zeta}_{x,r}(\,\cdot\,,\O)$, respectively.
However, we will use this notation here only in the case of homogeneous sets for which the choice of $r$ is unimportant, provided it is chosen sufficiently small; see Exercise~\ref{5.163/4} for an illustration of this statement.
See also, however, Remark~\ref{5.164/5} below.

\medskip

$(b)$~More generally, one could define the local fractal zeta function of an arbitrary relative fractal drum $(A,\O)$, where $A\subseteq\eR^N$, $\O\subseteq\eR^N$ (not necessarily open); see Remark~\ref{5.11/2}$(a)$ above.
To do so, it suffices to replace $B_r(x)$ by $A\cap B_r(x)$ in Definition~\ref{5.161/4} above.

\medskip

$(c)$~In some cases, in Definition~\ref{5.161/4}, it may be helpful to replace $B_r(x)$ by $A:=B_r(x)^c$, its complement in $\eR^N$. 
In that case, we have $A_t=B_{r-t}(x)^c$, for all $t\in(0,r)$.
(See the end of Exercise \ref{5.163/4} below for an illustration of this comment.)

\medskip

$(d)$ Another way to associate fractal zeta functions to unbounded subsets of $\eR^N$ is provided in the second author's forthcoming Ph.D.\ thesis~\cite{ra}, which extends (via `fractal zeta functions at infinity' and not via `local fractal zeta functions') the theory developed in~[LapRa\v Zu1--5]. 
\end{remark}

\medskip

\begin{exercise}\label{5.163/4}
We leave it to the interested reader to illustrate Definition~\ref{5.161/4} by calculating the local distance and tube zeta functions of $\eR^N$ (i.e., of $\O:=\eR^N$ in Definition~\ref{5.161/4}), as well as to deduce from this computation the (local) complex dimensions of $\eR^N$ (i.e., the poles of either $\zeta_x(\,\cdot\,,\eR^N)$ or of $\widetilde{\zeta}_x(\,\cdot\,,\eR^N)$).
As it turns out, in the present case, the local fractal zeta functions 
are independent of $x\in\eR^N$ (and are the same for $\zeta_x$ and $\widetilde{\zeta}_x$) and hence, so are the complex dimensions.
Furthermore, the complex dimensions are also independent of $r$, chosen sufficiently small.
One finds that the set of (local) complex dimensions of $\eR^N$, computed via the tube zeta function and denoted by $\dim_{loc}\eR^N$ since it is independent of $x\in\eR^N$, is given as follows:
$$
\dim_{loc}\eR^N=\{0,1,2,\ldots,N\},
$$
since $\widetilde{\zeta}_x$ in this case coincides with the tube zeta function of the $N$-dimensional ball $B_r(x)$; that is,
$$
\begin{aligned}
\widetilde{\zeta}_x(s,\eR^N)&=\int_0^\d t^{N-s-1}|B_{r+t}(x)\cap\O|\,\D t=\int_0^\d t^{N-s-1}\omega_N(r+t)^N\,\D t\\
&=\omega_N\sum_{k=0}^N \binom Nkr^k\int_0^{\d}t^{s-k-1}\,\D t=\omega_N\sum_{k=0}^N \binom Nk\frac{r^k\d^{s-k}}{s-k}.
\end{aligned}
$$
On the other hand, interestingly, since $\dim_BB_r(x)=N$, by Remark~\ref{2.71/2}$(a)$, the local dimensions of $\eR^N$ calculated via the local distance zeta function are given by
$$
\dim_{loc}'\eR^N=\{0,1,2,\ldots,N-1\},
$$
since by~\eqref{equ_tilde} we have
$$
\begin{aligned}
{\zeta}_x(s,\eR^N)&=\d^{s-N}|B_{r+\d}(x)|+(N-s)\,\widetilde{\zeta}_x(s,\eR^N)\\
&=\omega_N\left(\d^{s-N}(r+\d)^N+\sum_{k=0}^{N-1}\binom Nk\frac{r^k\d^{s-k}(N-s)}{s-k}-r^{N}\d^{s-N}\right).
\end{aligned}
$$

\medskip

\begin{xtolerant}{600}{}
If one replaces $B_r(x)$ by $B_r(x)^c$ in Definition~\ref{5.161/4} (as is suggested in part $(c)$ of Remark~\ref{5.161/2}), one obtains the same local complex dimensions.
(More details about this example are provided in \cite{fzf}.)
\end{xtolerant}
\end{exercise}

\medskip

\begin{remark}\label{5.164/5}
By considering inhomogeneous fractal sets $\O$, whose fractal dimension changes from point to point, and by suitably modifying the above definition of a local zeta function (Definition~\ref{5.161/4}), one could potentially develop a theory of multifractal zeta functions and their associated complex dimensions under hypotheses that are much more general than those made in earlier works on this subject; compare with~\cite{lapidusrock}, \cite{LaLeRo} and~\cite{ElLaMaRo} (as partly described in~\cite[Section~13.3]{lapidusfrank12}) and the relevant references therein (including~\cite{lemen}).
One could also adapt (in the obvious way) the above definition (Definition~\ref{5.161/4}) to an arbitrary set $\O\subseteq\eR^N$ equipped with a (suitable) measure $\eta$, under hypotheses that are significantly more general than for the `multifractal measures' considered in the above references and in the classic literature on this subject.
We close this discussion by noting that it would then be important to consider radii $r$ which are sufficiently small, or even to use a suitable limiting or averaging procedure in an appropriate modification of Definition~\ref{5.161/4}. 
\end{remark}

\section{Meromorphic extensions of the spectral zeta functions of fractal drums}\label{szf}

In this section, excerpted and adapted from~\cite[Section~4.3]{fzf}, we will review some of the known results concerning the spectral asymptotics of (relative) fractal drums,
with emphasis on the leading asymptotic behavior of the spectral counting function (or, equivalently, of the eigenvalues).
The corresponding remainder estimate, obtained by the first author in \cite{Lap1}, is expressed in terms of the upper box dimension of the boundary of the drum.
It turns out that the corresponding spectral zeta functions of these fractal drums have 
a (nontrivial) meromorphic extension; see Theorem~\ref{wb2} below. This fact was already observed by the first author in [Lap2--3] 
by other means, but also by using the error estimates of \cite{Lap1}.

\medskip

However, we give a slightly more precise formulation of the results here (and in~\cite{fzf}) and (at least in a useful special case) the proof of Theorem~\ref{wb2} provided in~\cite[Section~4.3]{fzf} is more elementary, in the sense that it does not rely on Tauberian (or Abelian) theory.
On the other hand, as was alluded to just above, the proof still relies (as in~[Lap2--3]) 
on the key error estimates obtained in~\cite{Lap1} (and recalled in a special case in Theorem~\ref{wb2c} below).

\medskip

We note that we can deduce from the above results an upper bound for the abscissa of meromorphic convergence of the spectral zeta function; see Theorem~\ref{wb2} below (which was already obtained in [Lap2--3]). 
We point out that by using in a crucial way some of the results of~\cite{fzf} and~\cite{mezf} (recalled in Theorem~\ref{ty} above), we can establish the optimality of the aforementioned upper bound; this is a new result, discussed in more details in~\cite[Section~4.3]{fzf}.

\medskip

We can now describe the main setting for this section.

\medskip

For a given relative fractal drum $(A,\O)$ in $\eR^N$,\footnote{In particular, this means that $|\O|<\ty$.} we are interested in 
the corresponding {\em Dirichlet eigenvalue problem}\label{dirichlet_ep}, 
defined on the  (possibly disconnected) open set
$\O_A:=\O\setminus \ov A$.\footnote{For example, if $\O$ is the unit square, and $A$ is the Sierpi\'nski carpet, then $\O_A$ is the union of a disjoint countable family of open squares.} In other words, we want to find all the ordered pairs $(\mu,u)\in\Ce\times H_0^1(\O_A)$ such that $u\ne0$ and
\begin{equation}\label{eig}
\left\{\begin{aligned}
-\DD u&=\mu u&&\mbox{ in $\O_A$,}\\ 
u&=0&&\mbox{ on $\pa(\O_A)$,}
\end{aligned}\right.
\end{equation}
in the variational sense (see, e.g., \cite{brezis}, \cite{LioMag}, \cite{Lap1} and the relevant references therein).
Here, $H_0^1(\O_A):=W_0^{1,2}(\O_A)$ is the standard {\em Sobolev space}
(see, e.g., \cite{brezis} and \cite{gt}),
 and  $\DD$ is the Laplace operator.%

\medskip

Throughout this section, we could assume equivalently that the relative fractal drum $(A,\O)$ is of the form of a standard fractal drum $(\pa\O_0,\O_0)$.
Indeed, it suffices to apply the quoted results (from \cite{Lap1}, for example), to the ordinary fractal drum $(\pa\O_0,\O_0)$, which is precisely what we will do,
implicitly.

\medskip

Recall the well-known fact that the (eigenvalue) spectrum of the Dirichlet eigenvalue problem 
(\ref{eig}) is discrete and consists of an infinite and 
 divergent sequence $(\mu_k)_{k\geq 1}$ of positive numbers
(called eigenvalues), without accumulation point (except $+\ty$) and which can be written in nondecreasing order according to multiplicity as follows:\footnote{Here, and thereafter, in order to avoid trivialities, we assume implicitly that all of the open sets $\O\subset\eR^N$ involved are nonempty.}
$$
0<\mu_1\le\mu_2\le\dots\le\mu_k\le\dots,\q \lim_{k\to\ty}\mu_k=+\ty.
$$
Furthermore, each of the eigenvalues $\mu_k$ is of finite multiplicity.
Moreover, if $\O_A$ is connected, then the first (or `principal') eigenvalue $\mu_1$ is of multiplicity one (i.e., $\mu_1<\mu_2$); see~\cite{gt}.
Because the Laplace operator is symmetric, each of its eigenvalues has equal algebraic and geometric multiplicities.
We will say for short that the sequence of eigenvalues $(\mu_k)_{k\geq 1}$ {\em corresponds} to the relative fractal drum $(A,\O)$.

\medskip

\begin{defn}\label{spectrum} We will define the {\em spectrum of a 
relative fractal drum} $(A,\O)$ in $\eR^N$, 
denoted by $\s(A,\O)$, as the sequence of the square roots of the eigenvalues of problem (\ref{eig}); that is,
\begin{equation}\label{eig1/2}
\s(A,\O):=(\mu_k^{1/2})_{k\ge1}.
\end{equation}
Physically, the values of $\mu_k^{1/2}$, $k\in\eN$, are interpreted as the (normalized)\footnote{When $N=1$, 
see \cite[Footnote 1 on page 2]{lapidusfrank12}.} {\em frequencies of 
the relative fractal drum}. The eigenvalues are scaled here with the exponent $1/2$, for technical (and physical) reasons.
\end{defn}

\medskip

\begin{defn}\label{spectral_zf}
The {\em spectral zeta function
$\zeta^*_{(A,\O)}$
 of a relative fractal drum} $(A,\O)$ in $\eR^N$ is given by\footnote{The spectral zeta 
function of a fractal string $\mathcal{L}=(l_j)_{j\ge1}$,
where $\mathcal{L}$ is viewed as a relative fractal drum $(A_{\mathcal{L}},\O_{\mathcal{L}})$, is given by $\zeta^*_{\mathcal{L}}(s)=\sum_{k,j=1}^\ty(k\cdot l_j^{-1})^{-s}=
\zeta(s)\zeta_{\mathcal{L}}(s)$, where $\zeta=\zeta_R$ is the classic Riemann zeta function \cite{Titch2} and $\zeta_{\mathcal{L}}$ is the 
geometric zeta function of $\mathcal{L}$;
see [Lap2--3], 
\cite{LaMa2} and \cite[Section 1.3]{lapidusfrank12}.}%
\begin{equation}\label{szeta}
\zeta^*_{(A,\O)}(s):=\sum_{k=1}^\ty\mu_k^{-s/2},
\end{equation}
for $s\in\Ce$ with $\re s$ sufficiently large.
\end{defn}

\medskip

The newly introduced definitions of the spectrum $\s(A,\O)$ and of the spectral zeta function $\zeta_{(A,\O)}^*$ of a relative fractal drum is in agreement with the definition of the spectrum of a bounded fractal string
$\mathcal{L}=(l_j)_{j\ge1}$ given in \cite[p.\ 2]{lapidusfrank12} or more generally, of a fractal drum (see, e.g.,
[{Lap1--3}]).
(See also \cite[Equation (1.45), p.\ 29]{lapidusfrank12} and \cite[Appendix B]{lapidusfrank12}, along with the relevant references therein, including~\cite{gil} and~\cite{sel1}.)
Note that the sequence 
\begin{equation}\label{eig-1}
\mathcal{L}(A,\O):=(\mu_k^{-1/2})_{k\geq 1},
\end{equation} 
which consists of the reciprocal frequencies in $\s(A,\O)$, is also a fractal string (possibly unbounded; that is, $\sum_{k=1}^{\ty}\mu_k^{-1/2}=\ty$). Obviously, the spectral zeta function of a relative fractal drum $(A,\O)$ is by definition equal to the geometric zeta function of the fractal string $\mathcal{L}(A,\O)$.
It is clear that $D(\zeta_{(A,\O)}^*)\ge0$.

\medskip

We stress that the usual definition of 
the spectrum involves the sequence of eigenvalues $(\mu_k)_{k\ge1}$ rather then the sequence of their 
square roots $(\mu_k^{1/2})_{k\ge1}$, as in Equation (\ref{eig1/2}).
As was already mentioned, we prefer the definition of the spectrum $\s(A,\O)$ given in Equation (\ref{eig1/2}) and hence, the use of the exponent $-s/2$ (rather than of $-s$) in the definition of the 
spectral zeta function $\zeta_{(A,\O)}^*$ in Equation (\ref{szeta}).
The reason is that in this case, Proposition \ref{spectral2} and Theorem \ref{wb2} below have a more elegant form. See [Lap2--3] 
and \cite[p.\ 29 and Appendix B]{lapidusfrank12} and compare, for example, with~\cite{gil} and~\cite{sel1}.

\medskip

The following theorem is a partial extension of \cite[Theorem 2.1]{lapidusfrank12} (or of the corresponding result for fractal sprays in [Lap2--3]) 
to the present context.

\medskip

\begin{theorem}\label{spectral_spray}
Let $(A_0,\O_0)$ be a base relative fractal drum in $\eR^N$, and let $\mathcal{L}=(\g_j)_{j\ge1}$ be a nonincreasing sequence of positive numbers tending to zero $($and repeated according to multiplicity$)$, i.e., a $($not necessarily bounded$)$ fractal string.
Assume that $(A_j,\O_j)$, $j\ge 1$, is a disjoint sequence of relative fractal drums, each of which is obtained by an isometry of $\g_j(A_0,\O_0)=(\g_j A_0,\g_j\O_0)$.
Let $(A,\O)=\bigcup_{j\ge1}(A_j,\O_j)$ be the corresponding relative fractal spray, generated by $(A_0,\O_0)$ and $\mathcal{L}$.
Then, assuming that $s\in\Ce$ is such that $\re s>\max\{D(\zeta^*_{(A_0,\O_0)}),\ov\dim_B\mathcal{L}\}$, we have
\begin{equation}\label{3.1.54}
\zeta^*_{(A,\O)}(s)=\zeta^*_{(A_0,\O_0)}(s)\cdot\zeta_{\mathcal{L}}(s);\footnote{By the principle of analytic continuation, 
Equation \eqref{3.1.54} continues to hold on any domain to which $\zeta_{\mathcal{L}}$ and $\zeta_{(A_0,\O_0)}^*$
can be meromorphically extended. (See, in particular, Theorem \ref{wb2} below.) A similar comment applies to
Equation \eqref{just_below}.}
\end{equation}
hence, for all $s\in\Ce$ with $\re s$ sufficiently large,
\begin{equation}
\zeta^*_{(A,\O)}(s)=\sum_{k=1}^\ty(\mu_k^{(0)})^{-s/2}\sum_{j=1}^\ty \g_j^s,
\end{equation}
where $(\mu_k^{(0)})_{k\ge1}$ is the sequence of eigenvalues corresponding to the relative fractal drum $(A_0,\O_0)$.
Furthermore,
\begin{equation}
D(\zeta^*_{(A,\O)})=\max\{D(\zeta^*_{(A_0,\O_0)}),\ov\dim_B\mathcal{L}\}.
\end{equation}
In particular, if $\g_j=\g^j$ for some fixed $\g\in(0,1)$, and each $\g^j$ is of multiplicity $b^j$, where $b\in\eN$, $b\ge2$, then for $\re s>D(\zeta_{(A_0,\O_0)}^*)$
\begin{equation}\label{just_below}
\zeta^*_{(A,\O)}(s)=\frac{b\g^s}{1-b\g^s}\sum_{k=1}^\ty(\mu_k^{(0)})^{-s/2}=\frac{b\lambda^s}{1-b\lambda^s}\zeta_{(A_0,\O_0)}^*(s),
\end{equation}
and
$$
D(\zeta^*_{(A,\O)})=\max\{D(\zeta^*_{(A_0,\O_0)}), \log_{1/\g}b\}.
$$
\end{theorem}

\medskip

\begin{remark}\label{3.1.601/2}
The factorization formula for the case of fractal sprays (and of fractal strings, in particular), was first observed in
[{Lap2--3}].
In the special case of fractal strings, it has proved to be very useful; see, especially,
[{Lap2--3, LapPo1--3, LapMa1--2, HeLap, Lap--vFr1--3, Tep1--2, LalLap1--2, HerLap1--5}].
See also, e.g., \cite[Sections~1.4 and~1.5]{lapidusfrank12} and \cite[Chapters~6,~9,~10 and~11]{lapidusfrank12}, both for the case of fractal strings and (possibly generalized or even virtual) fractal sprays.
\end{remark}

\medskip

Let $\O_0$ be a (nonempty) bounded open subset of $\eR^N$, and $\s(\pa\O_0,\O_0)=(\mu_k^{(0)})_{k\ge1}$ (that is, $(\mu_k^{(0)})_{k\ge1}$ is the sequence of eigenvalues of $-\DD$ with zero
(or Dirichlet) boundary data on
$\pa\O_0$, counting the multiplicities of the eigenvalues).
Then, the following classical asymptotic result holds, known as {\em Weyl's law}\label{weyl} 
[We1--2]:\footnote{Here, the symbol $\sim$ means that the ratio of the left and right sides of~\eqref{ch} tends to $1$ as $k\to\ty.$}
\begin{equation}\label{ch}
\mu_k^{(0)}\sim \frac{4\pi^2}{(\o_N|\O_0|)^{2/N}}\cdot k^{2/N}\q\mbox{as\qs$k\to\ty$,}
\end{equation}
where $\o_N=\pi^{N/2}/(N/2)!$ is the volume of the unit ball in $\eR^N$.\footnote{For odd $N$, we have $\left(\frac{N}{2}\right)!=\frac N2(\frac N2-1)\dots\frac12$, since $\left(\frac{N}{2}\right)!:=\Gamma(\frac N2+1)$,
where $\Gamma$ is the classic gamma function.}
The main result of this section is contained in Theorem \ref{wb2}. Its proof is based on the asymptotic result due to the first author, stated here in Theorem \ref{wb2c}, combined with 
Proposition \ref{spectral2}.

\medskip

In 1912, in~[We1--2], 
Hermann Weyl has obtained the asymptotic result stated in Equation (\ref{ch}) for piecewise smooth boundaries.
This result has since then been extended to a variety of settings
(for example, to smooth compact Riemannian manifolds with or without boundary, various boundary conditions, broader classes of elliptic operators, fractal boundaries, etc.).
See, for example, the well-known treatises by Courant and Hilbert
\cite[Section VI.4]{courant_hilbert} and by Reed and Simon \cite{resi}, along with \cite{ho3}, the introduction of~\cite{Lap1}, [Lap2--3] 
and \cite[Section 12.5 and Appendix B]{lapidusfrank12}.
It has been extended by G.\ M\'etivier in [M\'et1--3] 
during the 1970s to arbitrary bounded subsets of $\eR^N$ (in the present case of Dirichlet boundary conditions).
Furthermore, in this general setting (for example), sharp error estimates, expressed in terms of the upper Minkowski (or box) dimension of the boundary $\O_0$,
were obtained by the first author in the early 1990s in \cite{Lap1}; see Theorem \ref{wb2c} below, along with the comments following it for further extensions to
other boundary conditions and higher order elliptic operators (with possibly variable coefficients).

\medskip

In the following result, we consider a class of bounded open subsets $\O_0$ of $\eR^N,$ such that the corresponding sequence of eigenvalues $(\mu_k^{(0)})_{k\ge1}$ satisfies an asymptotic condition involving the second term as well:
\begin{equation}\label{ch2}
\mu_k^{(0)}= \frac{4\pi^2}{(\o_N|\O_0|)^{2/N}}\cdot k^{2/N}+O(k^{\gamma})\q\mbox{as\qs$k\to\ty$.}
\end{equation}
Here, we assume that $\gamma\in(-\infty,2/N)$.
It will also be convenient to write $\zeta_{\O_0}^*=\zeta_{(\pa\O_0,\O_0)}^*$, and more generally, 
$\zeta_{\O_0}^*=\zeta_{(A_0,\O_0)}^*$, provided $A_0$ and $\O_0$ are disjoint.
We say for short that $\zeta_{\O_0}^*$ is the {\em spectral zeta function of}\label{szfs} (the Dirichlet Laplacian on) $\O_0\subset\eR^N$.

\medskip

In the sequel, given a meromorphic (or, more generally, a complex-valued) function $f=f(s)$ (initially defined on some domain $U\subseteq\Ce$), we denote by $D_{mer}(f)$ the {\em abscissa of meromorphic convergence of} $f$.
By definition, $D_{mer}(f)\in\eR\cup\{\pm\ty\}$ is the unique extended real number such that $\{\re s>D_{mer}(f)\}$ is the {\em maximal} (i.e., largest) open right half-plane (of the form $\{\re s>\alpha\}$, for some $\alpha\in\eR\cup\{\pm\ty\}$) to which $f$ can be meromorphically extended.\footnote{By using the principle of analytic continuation, it is easy to check that this notion is well defined.}
Clearly, with the notation introduced earlier (in Section~\ref{disttube}) for $D_{hol}(f)$, the abscissa of holomorphic convergence of $f$, we always have $D_{mer}(f)\leq D_{hol}(f)$.

\medskip

\begin{prop}\label{spectral2}
Assume that $\O_0$ is an arbitrary bounded open subset of $\eR^N$ such that the corresponding sequence of eigenvalues of $-\DD$, with zero $($or Dirichlet$)$ boundary data on
$\pa\O_0$, counting the multiplicities of eigenvalues, satisfies the asymptotic condition \eqref{ch2}, where $C$ is a positive constant and $\gamma<2/N$.
Then the spectral zeta function 
\begin{equation}\label{84.1/2}
\zeta_{\O_0}^*(s)=\sum_{k=1}^\ty(\mu_k^{(0)})^{-s/2}
\end{equation}
possesses a unique meromorphic extension {\rm({\it at least})} to the open half-plane
\begin{equation}\label{spectral2a}
\big\{\re s>N-(2-\gamma N)\big\}.\footnote{As we see, the meromorphic extension vertical strip, to the left of the vertical line
 $\{\re s=N\}$, is of width at least $2-\gamma N$.}
\end{equation}
In other words, $D_{mer}(\zeta_{\O_0}^*)\le N-(2-\gamma N)$.
The only pole of $\zeta_{\O_0}^*$ in this half-plane is $s=N$, and in particular, $D(\zeta_{\O_0}^*)=N$. Furthermore, it is simple and
\begin{equation}
\res(\zeta_{\O_0}^*,N)=\frac{N\o_N}{(2\pi)^N}|\O_0|.
\end{equation}
\end{prop}

\medskip

Theorem \ref{spectral_spray}, combined with Proposition \ref{spectral2}, generalizes \cite[Theorem 1.19]{lapidusfrank12} to the $N$-dimensional case.
Now, we can state the main result of this section, which shows that the abscissa of meromorphic convergence of $\zeta_{\O_0}^*$ does not
exceed the upper box dimension of the boundary $\pa\O_0$ relative to $\O_0$, called the  dimension of $\pa\O_0$ in \cite{Lap1}.

\medskip
\begin{xtolerant}{500}{}
\begin{theorem}[{{[Lap2--3]}}]\label{wb2}
Let $\O_0$ be an arbitrary $($nonempty$)$ bounded open subset of $\eR^N$ such that $\ov\dim_B(\pa\O_0,\O_0)<N$. 
Then the spectral zeta function $\zeta_{\O_0}^*$ of $\O_0$
is holomorphic in the open half-plane $\{\re s>N\}$. Furthermore, $\zeta_{\O_0}^*$
can be $($uniquely$)$ meromorphically extended from $\{\re s>N\}$ to $\{\re s>\ov\dim_B(\pa\O_0,\O_0)\}$. In other words,
\begin{equation}\label{wb2mer}
D_{mer}(\zeta_{\O_0}^*)\le \ov\dim_B(\pa\O_0,\O_0).
\end{equation}
 Moreover, $s=N$ is the only pole of $\zeta_{\O_0}^*$ in the half-plane $\{\re s>\ov\dim_B(\pa\O_0,\O_0)\}$; it is simple and
\begin{equation}
\res(\zeta_{\O_0}^*,N)=\frac{N\o_N}{(2\pi)^N}|\O_0|.
\end{equation}
\end{theorem}
\end{xtolerant}

\medskip

Clearly, in light of the second part of Theorem~\ref{wb2}, it then follows that $D_{hol}(\zeta_{\O_0}^*)=N$.
Furthermore, since spectral zeta functions are given by standard Dirichlet series (with positive coefficients) and the Dirichlet Laplacian has infinitely many eigenvalues, we must also have $D_{hol}(\zeta_{\O_0}^*)=D(\zeta_{\O_0}^*)$ and hence, $D_{hol}(\zeta_{\O_0}^*)=D(\zeta_{\O_0}^*)=N$.
Therefore, the next corollary follows immediately.

\medskip

\begin{cor}
Under the assumptions of Theorem~\ref{wb2}, we have
\begin{equation}
D_{mer}(\zeta_{\O_0}^*)<D_{hol}(\zeta_{\O_0}^*)=D(\zeta_{\O_0}^*)=N.
\end{equation}
\end{cor}

\medskip

Theorem \ref{wb2} was already obtained by the first author in [Lap2--3].
We point out, however, that the statement of Theorem~\ref{wb2} given here is a little bit more precise than the one in [Lap2--3]. 
Furthermore, by using Proposition \ref{spectral2} above and Theorem~\ref{wb2c} below, and at least in a special case, a new (more elementary) proof of Theorem~\ref{wb2} can be given, which is not based on a Tauberian (or Abelian) argument (as in~[Lap2--3]). 
However, this proof still relies in an essential manner on the results of~\cite{Lap1} recalled in Theorem~\ref{wb2c}.
Alternatively, and in full generality, as is well known to the experts in spectral geometry and spectral theory, one can use a Tauberian-type theorem~\cite{Pos} (really, an Abelian-type argument, in the sense of~\cite{Sim}, for example) to deduce Theorem~\ref{wb2} from Theorem~\ref{wb2c} below or, more directly, from the equivalent form of Theorem~\ref{wb2c} which is stated in Remark~\ref{footnote34}; see, e.g., \cite[Appendix~A]{Lap1}, \cite{Sim} and [Lap2--3]. 
Indeed, the spectral zeta function $\zeta^*_{\O_0}$ is essentially equal to the Mellin transform of the eigenvalue (or spectral) counting function $N_{\nu}$.\footnote{Let $(\mu_k^{(0)})_{k\ge1}$ be the sequence of eigenvalues of $-\DD$, where $\DD$ is the Dirichlet Laplacian associated with a given bounded open subset $\O_0$ of $\eR^N$.
We denote the {\em eigenvalue counting function of the fractal drum} by $N_\nu$; i.e.,
\begin{equation}\label{Nnu}\nonumber
N_\nu(\mu):=\#\{k\in\eN:\mu_k^{(0)}\le\mu\},\q\mbox{for\q$\mu>0$,}
\end{equation}
with the multiplicities of the eigenvalues being taken into account.
It is also called the {\em spectral counting function} in the literature; see, e.g., [Lap1--5], 
[Lap--vFr1--3] 
and the relevant references therein.}

\medskip

We also note here that an equivalent form of Theorem~\ref{wb2c} below, stated in terms of the eigenvalue counting function, provides a partial resolution of the 
{\em modified Weyl--Berry\label{wbconj} conjecture}.
See \cite[Corollary 2.1]{Lap1}, as well as \cite[Theorems 2.1 and 2.3]{Lap1}, along with the comments following Theorem~\ref{wb2}, 
for a more general statement involving positive uniformly elliptic linear differential operators 
(with variable and possibly nonsmooth coefficients) and mixed Dirichlet--Neumann boundary conditions.

\medskip

\begin{theorem}[{{[{Lap1}]}}]\label{6.7}
\label{wb2c}
Let $\O_0$ be an arbitrary $($nonempty$)$ bounded open subset of $\eR^N$. As before, we let $\widetilde D:=\ov\dim_B(\pa\O_0,\O_0)$
and let $(\mu_k^{(0)})_{k\ge1}$ be the sequence of eigenvalues of $-\DD$, where $\DD$ is the Dirichlet Laplacian on $\O_0$.
Then the following conclusions hold$:$

\medskip

$(i)$ If $\widetilde D\in(N-1,N]$, then for any $d>\widetilde D$,
\begin{equation}\label{wb2c_1}
\mu_k^{(0)}=\frac{4\pi^2}{(\o_N|\O_0|)^{2/N}}\cdot k^{2/N}+O(k^{(2+d-N)/N})\q\mathrm{as}\qs k\to\ty.
\end{equation}

\medskip

$(ii)$ If $\widetilde D=N-1$, then for any $d>\widetilde D$,
\begin{equation}\label{wb2c_2}
\mu_k^{(0)}=\frac{4\pi^2}{(\o_N|\O_0|)^{2/N}}\cdot k^{2/N}+O(k^{(2+d-N)/N}\log k)\q\mathrm{as}\qs k\to\ty.
\end{equation}

\medskip

Furthermore, in each of the cases $(i)$ and $(ii)$, the choice of $d=\widetilde D$ is allowed, provided $\M^{*\widetilde D}(\pa\O_0,\O_0)<\ty$.
\end{theorem}

\medskip

\begin{remark}\label{footnote34}
Via arguments that are standard in spectral theory and spectral geometry, Equation~\eqref{wb2c_1} is shown to be equivalent to
$$
N_\nu(\mu)=(2\pi)^{-N}\o_N|\O_0|\cdot\mu^{N/2}+O(\mu^{d/2})\q\mbox{as\qs$\mu\to+\ty$},
$$
where $N_\nu$ is the eigenvalue counting function of the fractal drum. A similar remark holds also for Equation~\eqref{wb2c_2}; one must simply replace $O(\mu^{d/2})$ with $O(\mu^{d/2} \log \mu)$ on the right-hand side of the above remainder estimate.
\end{remark}

\medskip

\begin{xtolerant}{1000}{}
Various aspects of the modified Weyl--Berry conjecture are studied in [{Berr1--2}], 
\cite{BroCar}, 
[{Lap1--3}], 
[{LapPo1--3}], 
\cite{Cae}, \cite{vBGi}, \cite{HamLa}, \cite{FlVa}, \cite{Ge}, \cite{GeSc}, \cite{MoVa} and [Lap--vFr1--3] 
among other references.
An early survey of this subject is provided in~\cite{Lap3} as well as in the introduction of~\cite{Lap1}, while a short, but more recent, survey of this subject along with additional relevant references is provided in~\cite[Subsection~12.5.1]{lapidusfrank12}.
(See also [Lap7--8].)
The eigenvalue counting function form of the result stated in case $(ii)$ of Theorem \ref{wb2c}, that is, in the nonfractal case where $\widetilde D=N-1$, and under the additional assumption
that $\mathcal{M}^{*(N-1)}(\pa\O)$ is finite, was already obtained in M\'etivier's\label{mt} work \cite[Theorem 6.1 on page 191]{mt}.\footnote{The special case of the corresponding error estimate for domains with piecewise smooth boundaries was obtained early on by Courant (as described in~\cite{courant_hilbert}), while the extension to smooth, compact Riemannian manifolds (with or without boundaries) was studied successively by H\" ormander~\cite{ho1} (see also~[H\"o2--3]), 
Seeley~[See2--3] 
and Pham The Lai~\cite{ph}, in particular.}
We note that M\'etivier stated his result without the explicit use
of box (that is, Minkowski) dimension or Minkowski content.
Results concerning
 the partition 
function (the trace of the heat semigroup) of the Dirichlet Laplacian
have been obtained 
by Brossard and Carmona in~\cite{BroCar}.
However, the main remainder estimate in \cite{BroCar} is now a consequence of the
results of \cite{Lap1} stated in Theorem~\ref{6.7}, whereas the converse is not true.
Indeed, as is well known, beyond the leading term, the spectral asymptotics for the trace of the heat semigroup do not imply the corresponding asymptotics for the eigenvalue counting function (or, equivalently, for the eigenvalues themselves).
\end{xtolerant}

\medskip


Although we have only discussed the Dirichlet Laplacian $-\DD$ thus far, we can also discuss the Neumann Laplacian,\footnote{Since the boundary $\partial\O_0$ is allowed to be fractal, the corresponding eigenvalue problem $-\DD u=\mu u$ in $\O_0$, with $\pa u/\pa n=0$ on $\pa\O_0$ (where $\pa u/\pa n$ stands for the normal derivative of $u$ along $\pa\O_0$) should, of course, be interpreted in the variational sense (that is, with $H_0^1(\O_0)$ replaced by the Sobolev space $H^1(\O_0):=W^{1,2}(\O_0)$); see, e.g.,~\cite{brezis}, \cite{LioMag} and~\cite{Lap1}, Section~4, especially, Subsection~4.2.B.} or general positive uniformly elliptic linear differential operators (with variable and possibly nonsmooth coefficients) of the form 
$$
\mathcal{A}
=\sum_{|\a|\le m,|\b|\le m}(-1)^{|\a|}D^{\a}(a_{\a\b}(x)D^{\b}),
$$
of order $2m$, described in \cite[Section 2.2]{Lap1}. 
All of these extensions of Theorem \ref{6.7} (and of its equivalent form given in Remark \ref{footnote34}) are also obtained in \cite{Lap1}; see Theorem 2.1 and its corollaries in \cite{Lap1}.
In the latter case, 
the assumed asymptotic expansion of the eigenvalues of $\mathcal{A}$ in the corresponding version of Proposition \ref{spectral2},
and implied by (or, actually, equivalent to) \cite[Theorem 2.1]{Lap1}, should be replaced by 
\begin{equation}\label{mu_m}
\mu_k^{(0)}=(\mu'_{\mathcal A}(\O_0))^{-2m/N}\cdot k^{2m/N}+O(k^\gamma),\q\mbox{as\qs$k\to\ty$},
\end{equation}
where $\mu'_{\mathcal{A}}(\O_0)$ is the ``Browder--G\aa rding measure'' of $\O_0$ defined, for example, in \cite{ho3} or in \cite[Equation $(2.18a)$ in Section 2.2]{Lap1}
in terms of the (positive definite, unbounded) quadratic form associated with $\mathcal{A}$,\footnote{More precisely,
$$
\mu'_{\mathcal{A}}(\O_0)=\int_{\O_0}\mu'_{\mathcal A}(x)\,\mathrm{d}x,
$$
where the ``Browder--G\aa rding density'' $\mu'_{\mathcal A}(x)$ is given (for a.e.\ $x\in\O_0$) by
$$
\mu'_{\mathcal A}(x):=(2\pi)^{-N}|\{\xi\in\eR^N:a'(x,\xi)<1\}|,
$$
with
$
a'(x,\xi)=\sum_{|\alpha|=|\beta|=m}a_{\alpha\beta}(x)\xi^{\alpha+\beta}
$
being the leading symbol of the nonnegative quadratic form associated with $\mathcal A$ and with $\xi^{\kappa}:=\xi_1^{\kappa_1}\cdots\xi_N^{\kappa_N}$ for $\xi\in\eR^N$ and $\kappa=(\kappa_1,\ldots,\kappa_N)\in(\eN\cup\{0\})^N$, as well as for a.e.\ $x\in\O_0$.
Note that, as a result, $\mu'_{\mathcal A}(\O_0)$ can be interpreted as a volume in phase space $\eR^{2N}\simeq\eR^N\times\eR^N$ (which, in the special case where $\mathcal A$ is a Schr\" odinger-type operator, is in agreement with the usual semiclassical limit of quantum mechanics; see, e.g., \cite{resi} and~\cite{Sim}).} $\gamma:=(2m+d-N)/N$, with $d\in(\widetilde{D},N)$ arbitrary and (as before) $\widetilde{D}:=\ov{\dim}_B(\partial\O_0,\O_0)$, the inner Minkowski dimension of $\O_0$ (i.e., the upper Minkowski dimension of the relative fractal drum $(\partial\O_0,\O_0)$).
Furthermore, we may take $d=\widetilde{D}$ provided $\mathcal{M}^{*\widetilde{D}}(\partial\O_0,\O_0)<\ty$.
We note that the remainder estimate~\eqref{mu_m} holds in the above form in the `fractal case' where $\widetilde{D}>N-1$ (or, equivalently, where $\widetilde{D}\in(N-1,N]$ since we always have $\widetilde{D}\in[N-1,N]$).\footnote{Here and in the sequel, we should replace $\widetilde{D}$ by $D$, where $D:=\ov{\dim}_B(\partial\O_0)$, the upper Minkowski dimension of the boundary $\partial\O_0$ in the case of Neumann (or, more generally, mixed Dirichlet--Neumann) boundary conditions.}

\medskip

We have just stated, in estimate~\eqref{mu_m} and the text following it, the analog (obtained in~\cite{Lap1}) of case $(i)$ of Theorem~\ref{6.7} above.
(Note that when $m=1$, estimate~\eqref{mu_m} reduces to estimate~\eqref{wb2c_1}.)
In the `nonfractal case' (or `least fractal case', still following the terminology of~\cite{Lap1}) where $\widetilde{D}=N-1$, the exact analog of part $(ii)$ of Theorem~\ref{6.7} also holds.
Namely, still according to (\cite{Lap1}, Theorem~2.1 and its corollaries), the precise counterpart of estimate~\eqref{mu_m} holds, with $O(k^{\gamma}\log k)$ (instead of $O(k^{\gamma})$) and with the same value of $\gamma$ as above, exactly as in estimate~\eqref{wb2c_2} of case $(ii)$ of Theorem~\ref{6.7} (which corresponds to the case when $m=1$).\footnote{With the notation introduced in Remark~\ref{footnote34} for the eigenvalue counting function $N_{\nu,\mathcal{A}}:=N_{\nu}$ (of the operator $\mathcal{A}$), estimate~\eqref{mu_m}  (and its counterpart when $\widetilde{D}=N-1$) can be written equivalently as follows:
$$
N_{\nu,\mathcal{A}}(\mu)=\mu_{\mathcal{A}}'(\O_0)\mu^{N/2m}+R(\mu),
$$
where $R(\mu):=O(\mu^{d/2m})$ in the fractal case when $\widetilde{D}>N-1$ and $R(\mu):=O(\mu^{d/2m}\log\mu)$ in the nonfractal case where $\widetilde{D}=N-1$.
(And, analogously, with $D=\ov{\dim}_B(\partial\O_0)$ instead of $\widetilde{D}=\ov{\dim}_B(\partial\O_0,\O_0)$ for Neumann or, more generally, mixed Dirichlet--Neumann boundary conditions.)}

\medskip

We now consider the consequences of the above error estimates for the spectral zeta function $\zeta_{\mathcal{A},\O_0}^{*}:=\zeta_{\O_0}^*$, defined (for $s\in\Ce$ with $\re s$ sufficiently large) by 
\begin{equation}\label{spectr_zeta_m}
\zeta_{\O_0}^*(s):=\sum_{k=1}^{\ty}(\mu_k^{(0)})^{-s/2m},
\end{equation}
via the sequence $(\mu_k^{(0)})_{k\geq 1}$, of eigenvalues of $\mathcal{A}$ (see the precise definition of the spectrum and the domain of the operator $\mathcal{A}$ 
given in \cite[Section 2.2]{Lap1}; see also \cite{LioMag} or \cite{mt}).

\medskip

Let us now assume that $\widetilde{D}<N$, in order for the analog of Weyl's asymptotic estimate to hold (in light of~\eqref{mu_m}).
It then follows from the above discussion (that is, from estimate~\eqref{mu_m} when $\widetilde{D}>N-1$ or from its counterpart when $\widetilde{D}=N-1$) that $\zeta_{\O_0}^*$ is holomorphic in the open half-plane $\{\re s>N\}$ and can be (uniquely) meromorphically extended to the strictly larger open half-plane $\{\re s>\widetilde{D}\}$, with a single (simple) pole in that half-plane, located at $s=N$ and of residue given by
\begin{equation}
\res\left(\zeta_{\O_0}^*,N\right)={N}\mu_{\mathcal A}'(\O_0).
\end{equation}
(This is true for any value $\widetilde{D}$ in $[N-1,N)$.\footnote{In light of~\eqref{mu_m} or its counterpart when $\widetilde{D}=N-1$, these results are established either by combining Theorem~\ref{spectral_spray} and a suitable analog of Proposition~\ref{spectral2} or else (as was first observed in~\cite{Lap3}) by using a Tauberian (or rather, Abelian) argument.})
Consequently, we deduce that the abscissa of holomorphic convergence of $\zeta_{\O_0}^*$ satisfies
\begin{equation}\label{831/2}
D_{hol}(\zeta_{\O_0}^*)={N},
\end{equation}
 whereas the abscissa of meromorphic convergence of $\zeta_{\O_0}^*$ satisfies the following inequality (which, in the special case when $m=1$, formally looks exactly like inequality~\eqref{wb2mer}):
\begin{equation}\label{zeta*mer}
D_{mer}(\zeta_{\O_0}^*)\leq{\widetilde{D}}.
\end{equation}
In particular (since $\widetilde{D}<N$, by assumption), we have that $D_{mer}(\zeta_{\O_0}^*)<D_{hol}(\zeta_{\O_0}^*)$.


\medskip

As is noted in [Lap2--3], 
all of these results rely on the analog of 
Theorem~\ref{wb2c} corresponding to uniformly elliptic differential operators $\mathcal{A}$ of order $2m$, 
which is obtained in \cite[Theorem 2.1 and Corollary 2.2]{Lap1}; see~[Lap2--3].

\medskip

The remainder estimates obtained in \cite{Lap1} are the best possible, in general, in the most important case of a fractal drum for which $N>\widetilde D>N-1$, $\M^{*\widetilde{D}}(\partial\O_0,\O_0)<\ty$ (and hence, $d:=\widetilde{D}$), and $\mathcal A$ is the Dirichlet Laplacian;\footnote{An easy verification shows that for this same family of examples, the analogous statement is true for the Neumann Laplacian, provided $\widetilde{D}$ is replaced by $D=\ov{\dim}_B(\partial\O_0)$ and $\M^{*\widetilde{D}}(\partial\O_0,\O_0)$ is replaced by $\M^{*{D}}(\partial\O_0)$ (and the eigenvalue $0$ is excluded from the definition of $\zeta_{\O_0}^{*}$ in Equation~\eqref{84.1/2}).} that is, in case $(i)$ of Theorem \ref{wb2c}, the stated error estimate (\ref{wb2c_1}) is sharp.
Indeed, there exists a one-parameter family of Minkowski measurable examples for which $\widetilde D$ takes all possible values in the allowed open interval $(N-1,N)$ 
and the error estimate (\ref{wb2c_1})
is sharp.\footnote{Recall from \cite[Corollary 3.2]{Lap1} that (since $\O_0$ is open and bounded) $\widetilde D:=\ov\dim_B(\pa\O_0,\O_0)$ always satisfies the following inequality: $N-1\le\widetilde D\le N$.\label{footnote44}}
(See \cite[Example 5.1 and Example 5.1']{Lap1}.)

\medskip

The aforementioned family of examples provided in~\cite{Lap1} does {\em not} consist of {\em connected} open sets (but each of the examples is such that $\partial\O_0$ is Minkowski measurable).
One may slightly modify these examples by appropriately opening small gates in each of the `teeth' of the `fractal combs' constructed in~\cite{Lap1}, much as in~\cite{BroCar} and~\cite{FlVa}.
For the resulting (simply connected) domains, the remainder estimates~\eqref{wb2c_1} (with $d:=\widetilde{D}$, because we have $\mathcal{M}^{*\widetilde{D}}(\partial\O_0,\O_0)<\ty$)\footnote{In fact, one can arrange for $\partial\O_0$ to remain Minkowski measurable.} are still sharp for any value of $\widetilde{D}\in(N-1,N)$.

\medskip

Still for the Dirichlet Laplacian, the same family of examples in~\cite[Example~5.1 and~5.1']{Lap1} (or its modification discussed just above) does not allow us to show that the inequality~\eqref{zeta*mer} is sharp.
Indeed, for those examples, using a result obtained in~\cite[Theorem~6.21]{lapidusfrank12}, one can prove that $D_{mer}(\zeta_{\O_0}^*)=-\ty<\widetilde{D}$.
(See~\cite{fzf}.)

\medskip

\begin{xtolerant}{1000}{}
However, using in an essential manner some of the results of~\cite[Sections~4.5 and~4.6]{fzf} partially described in Section~\ref{relative} above (especially in Theorem~\ref{ty} and Corollary \ref{5.171/2}), one can construct an example of a bounded open set $\O_0$ of $\eR^N$ (with $N\geq 1$ arbitrary), viewed as a relative fractal drum $(\partial\O_0,\O_0)$, and such that the inequality~\eqref{wb2mer} is sharp; that is, such that
\begin{equation}\label{841/2}
D_{mer}(\zeta_{\O_0}^*)=\ov{\dim}_B(\partial\O_0,\O_0)=:\widetilde{D}
\end{equation}
and hence, $\{\re s>\widetilde{D}\}$ is the largest open right half-plane to which $\zeta_{\O_0}^*$ can be extended.
(See~\cite[Section~4.3]{fzf} for more details.)
In fact, much as in Theorem~\ref{ty}, the boundary $\partial\O_0$ is both maximally hyperfractal and $\ty$-quasiperiodic.
(A simple modification enables us to show that the same is true for the Neumann Laplacian, except for $\widetilde{D}$ replaced by $D=\ov{\dim}_B(\partial\O_0)$.)
Here, $N-1<D<N$ and $\O_0=V_0\times(0,1)^{N-1}$, where $V_0$ is a geometric realization of the fractal string $\mathcal L$ provided by Theorem~\ref{ty}; see Corollary \ref{5.171/2}.
(Recall that $\mathcal L=(V_0,\partial V_0)$ is $\ty$-quasiperiodic and maximally hyperfractal; hence, so is $(\partial\O_0,\O_0)$.)
It follows that $\widetilde{D}=\ov{\dim}_B(\partial V_0,V_0)+N-1\in(N-1,N)$ and that each point of the critical line $\{\re s=\widetilde{D}\}$ is a singularity of $\zeta_{\O_0}^*$, so that $\zeta_{\O_0}^*$ cannot be meromorphically extended to the left of $\{\re s=\widetilde{D}\}$.
Therefore, in light of inequality~\eqref{wb2mer} of Theorem~\ref{wb2}, we deduce that~\eqref{841/2} holds, i.e., that inequality~\eqref{zeta*mer} is actually an equality for this example.
It then also follows that the abscissa of meromorphic convergence of the spectral zeta function of $\O_0$ as well as the abscissae of meromorphic, holomorphic and (absolute) convergence of the fractal (i.e., distance and tube) zeta functions of the relative fractal drum $(\partial\O_0,\O_0)$ all coincide with the (relative) upper Minkowski dimension $\widetilde{D}=\ov{\dim}_B(\partial\O_0,\O_0)$:\footnote{An entirely analogous result holds for the Neumann Laplacian (instead of the Dirichlet Laplacian) on the same open set $\O_0$, and with $\O_0$ instead of the relative fractal drum $(\partial\O_0,\O_0)$; in particular, as usual, $\widetilde{D}=\ov{\dim}_B(\partial\O_0,\O_0)$ is replaced by $D=\ov{\dim}_B(\partial\O_0)$.
(For the definition of $\zeta_{\O_0}^*$, one must also exclude the eigenvalue zero.)}
\begin{equation}\label{841/2+e}
D_{mer}(\zeta_{\O_0}^*)=\ov{\dim}_B(\partial\O_0,\O_0)=\widetilde{D}=D_{mer}(f)=D_{hol}(f)=D(f),
\end{equation}
for all $f\in\{\zeta_{\partial\O_0}(\,\cdot\,,\O_0),\widetilde{\zeta}_{\partial\O_0}(\,\cdot\,,\O_0)\}$.
\end{xtolerant}


\medskip

We leave it to the interested reader to investigate whether the above example can be modified (in the spirit of the previous discussion concerning the sharpness of the remainder estimates of~\cite{Lap1} recalled in part $(i)$ of Theorem~\ref{6.7}) so as to be connected (or even, simply connected) and, more generally, to be replaced by a one-parameter family of connected (or even, simply connected) bounded open subsets of $\eR^N$ such that inequality~\eqref{wb2mer} is sharp and $\widetilde{D}$ ranges through all of $(N-1,N)$ as the parameter varies.

\medskip

Even more generally, we can ask an entirely similar question, but now with inequality~\eqref{wb2mer} replaced by inequality~\eqref{zeta*mer}, now corresponding to elliptic operators of order $2m$ ($m\geq 1$ arbitrary) with variable coefficients and for Dirichlet or Neumann (or, more generally, mixed Dirichlet--Neumann boundary) conditions.
We expect the answer to be affirmative in this more general situation as well.
In particular, we expect inequality~\eqref{zeta*mer} to be sharp, in general, and therefore, to find examples for which $D_{mer}(\zeta_{\O_0}^*)=\widetilde{D}$ (resp., $=D$), where $\widetilde{D}=\ov{\dim}_B(\partial\O_0,\O_0)$ (resp., $D=\ov{\dim}_B(\partial\O_0)$), in the case of Dirichlet (resp., Neumann, or more generally, mixed Dirichlet--Neumann) boundary conditions.

\medskip

In the case of Neumann, or more generally, of mixed Dirichlet--Neumann boundary conditions, it follows from the results of \cite{Lap1} (and [Lap2--3]) 
that Theorem \ref{wb2c}, and hence also Theorem \ref{wb2} still hold (along with their more general counterparts for positive uniformly elliptic operators of order $2m$) 
provided that $\O_0$ is assumed to be a bounded open set of $\eR^N$ satisfying the {\em extension property} and $\widetilde D=\ov\dim_B(\pa\O_0,\O_0)$ (the upper, inner Minkowski dimension of $\pa\O_0$) is replaced by $D=\ov\dim_B(\pa\O_0)$, the upper Minkowski (or box) dimension of $\pa\O_0$
in the statement of Theorem \ref{wb2c} and Theorem \ref{wb2}.\footnote{It is clear from the definitions that $\widetilde D\le D$, and hence, it follows from footnote~\ref{footnote44} that $N-1\le\widetilde D\le D\le N$. Furthermore, there are natural examples of planar domains for which $\widetilde D<D$; see \cite[Note added in proof, p.\ 525]{Lap1} 
and the relevant reference therein, \cite{Tri}.}
(See, in particular, \cite[Theorem 2.3 and Corollary 2.2]{Lap1}.)
Recall that, by definition, the extension property states that every function in the Sobolev space $H^1(\O_0):=W^{1,2}(\O_0)$ can be extended to a function 
in $H^1(\eR^N):=W^{1,2}(\eR^N)$, and the resulting extension operator is a bounded linear operator from $H^1(\O_0)$ to $H^1(\eR^N)$. For example, a bounded domain 
$\O_0$ in $\eR^N$ satisfies the extension property if its boundary $\pa\O_0$ is of class $C^1$; see, e.g., \cite[Th\'eor\`eme IX.7]{brezis}. Note that, in this
latter case, $\dim_B(\pa\O_0,\O_0)=N-1$.

\medskip

Alternatively, the above mentioned results of \cite{Lap1} imply that
(still for Neumann or, more generally, mixed Dirichlet--Neumann boundary conditions)
 instead of satisfying the extension property, $\O_0$ can be assumed to satisfy the so-called
$(C')$-{\em condition} \cite[Definition 2.2]{Lap1}
(which is satisfied, for example, if $\O_0$ is locally Lipschitz, or satisfies either a `segment condition',
a `cone condition', or else is an open set with cusp; see \cite{mt} or \cite[Examples 2.1 and 2.2]{Lap1}), in which case we are necessarily in 
case (ii) of the counterpart of Theorem \ref{wb2c}, with $D\,\,(:=\ov\dim_B(\pa\O_0,\O_0))=N-1$.

\medskip

Recall that (as is proved by Jones in \cite{jones} and discussed in \cite[Example 4.2]{Lap1}, see also~\cite{mazja}) in two dimensions (i.e., when $N=2$), a simply connected domain $\O_0$
satisfies the 
{\em extension property}
(or is an {\em extension domain}) if and only if it is a {\em quasidisk} (i.e., a Jordan curve which is the quasiconformal image of the unit disk in $\eR^2$).
The boundary $\pa\O_0$ of a quasidisk is called a {\em quasicircle}, and the property of being a quasicircle can be characterized geometrically by a
{\em chord-arc condition}. Furthermore, a quasicircle can have any dimension between $1$ and $2$. See \cite{mazja} and \cite{pommerenke},
along with the relevant references therein, for a detailed discussion of quasidisks, quasicircles and extension domains. 
The class of quasicircles includes the classic Koch snowflake curve and its natural generalizations, as well as the Julia sets associated with the quadratic maps $z\mapsto z^2+c$ ($z\in\Ce$),
provided the parameter $c\in\Ce$ is sufficiently small. 
(Such Julia sets are necessarily connected because the corresponding complex parameter $c$ belongs to the main cardioid of the Mandelbrot set.)
Therefore, the Koch snowflake domain (and its generalizations) and the bounded domains having for boundary
the aforementioned Julia sets, are natural examples of quasidisks and hence, of extension domains.

\medskip

In higher dimensions, {\em extension domains} (i.e., domains of $\eR^N$ satisfying the (Sobolev) extension property) are more difficult to characterize.
However, it has been shown by Hajlasz, Koskela 
and Tuominen in [HajKosTu1--2] 
that a bounded domain $\O_0\st\eR^N$ is an extension domain if and only if it satisfies a certain functional
analytic condition
 and the {\em measure density condition};%
\footnote{The set $\O_0\stq\eR^N$ is said to satisfy the {\em measure density condition}
 (or to be a {\em lower Ahlfors regular $N$-set}) if there exists a positive constant $M$ such that
$$
|\O_0\cap B_r(x)|\le Mr^N,
$$%
for all $x\in\O_0$ and all $0<r\le1$, where $B_r(x)$ denotes the open ball of center $x$ and radius $r$ in $\eR^N$; see \cite{hajkotu1}.} see \cite[Theo\-rem~5]{hajkotu1}.

\medskip

Finally, we note that for Neumann boundary conditions, the above results concerning spectral asymptotics and spectral zeta functions 
also extend to higher-order
uniformly elliptic self-adjoint operators (with variable coefficients); see \cite{Lap1}.

\section{Classification of bounded fractal sets in Euclidean spaces}\label{class}

The classification of bounded subsets $A$ of Euclidean spaces, introduced in \cite{fzf}, is based on the asymptotic properties of the associated {\em tube functions} $t\mapsto|A_t|$ as $t\to0^+$.
It is a result of the development of the theory of bounded fractal strings and the associated complex dimensions from the early 1990s, as well as of the recent advances
in the theory of fractal zeta functions and complex dimensions of bounded sets in Euclidean spaces described in \cite{fzf}. 

\medskip

In the sequel, $A$ is any (nonempty) bounded subset of $\eR^N$, where $N$ is an arbitrary positive integer.
We begin with the roughest classification.

\medskip

$(a)$ $A$ is {\em Minkowski nondegenerate} (or simply {\em nondegenerate}), 
if there exists $D\ge0$ such that $0<\M_*^D(A)\le\M^{*D}(A)<\ty$. In particular, we then have $D=\dim_BA$.

\medskip

$(b)$ $A$ is a {\em Minkowski degenerate set} (or simply {\em degenerate}) if 
\medskip

     $\bullet$ either $D=\dim_BA$ exists and at least one of the corresponding $D$-dimensional Minkowski contents is degenerate (in other words, $\M_*^D(A)=0$ or $\M^{*D}(A)=+\ty$)
		
		\medskip
		
     $\bullet$ or else $\underline\dim_BA<\ov\dim_BA$.
		
		\medskip
		

We now introduce a finer classification of bounded sets in $\eR^N$, based on the asymptotic behavior of their tube functions.
First, we consider the case of Minkowski nondegenerate sets $A$. This is equivalent to saying that the tube function $t\mapsto |A_t|$
has the following form:
\begin{equation}\label{F}
|A_t|=t^{N-D}(F(t)+o(1))\quad\mbox{as\qs$t\to0^+$,}
\end{equation}
where (for some fixed, but sufficiently small, $\delta>0$) the range of the function $F:(0,\delta)\to\eR$ is bounded away from zero and infinity; that is, $0<\inf F\le\sup F<\ty$. Clearly, it then follows that
$$
\liminf_{t\to0^+} F(t)=\M_*^D(A),\quad \limsup_{t\to0^+} F(t)=\M^{*D}(A).
$$
 The idea underlying this classification is to introduce function-theoretic notions for bounded (or, equivalently, compact)  sets.\footnote{Note that by replacing $A$ by its closure $\ov{A}$, the (lower, upper) Minkowski dimension and content remain unchanged, along with $|A_t|$.
Hence, without loss of generality, we may assume that $A$ is compact instead of merely being bounded.}
More precisely, various properties of $A$ will be expressed in terms of the properties of an associated function~$F$ in the corresponding
tube formula~(\ref{F}).

\medskip

We shall need an auxiliary function $\rho=\rho(t)$, defined for $t>0$ small enough, such that

\begin{equation}\label{rho}
\mbox{$\rho=\rho(t)$ is decreasing, positive, continuous, and $\lim_{t\to0^{+}}\rho(t)=+\ty$.}
\end{equation}
\bigskip

\noindent
Let $A$ be a Minkowski nondegenerate bounded (or, equivalently, compact) subset of $\eR^N$. We say that

\medskip

     $\bullet$ $A$ is a {\em constant set},\label{constant} or a {\em Minkowski measurable set}, if there exists a finite and positive constant $\mathcal{M}$ such that (\ref{F}) is satisfied with $F(t)\equiv\mathcal{M}$.
     It then follows that $A$ is Minkowski measurable with Minkowski content $\mathcal{M}$.
		
		\medskip
		
     $\bullet$ $A$ is a {\em nonconstant set}\label{nonconstant} if there is no positive constant function $F$ satisfying (\ref{F}).

\bigskip

\noindent
We now classify Minkowski nondegenerate sets that are not constant, i.e., sets that are not Minkowski measurable.
Let $A$ be a nonconstant (Minkowski nonmeasurable) set in $\eR^N$.

\medskip

     $\bullet$ $A$ is a {\em periodic set}\label{periodic_set} if (\ref{F}) holds with $F$ of the form  $F(t)=G(\rho(t))$ for all positive $t$ small enough, where $G$ is a 
		periodic function
     and $\rho$ satisfies conditions (\ref{rho}). In the applications, we often have $\rho(t)=\log1/t$, like in the case of the Cantor set 
     or of the Sierpi\'nski carpet. 
		 The value of the minimal period of $G$ is called the {\em oscillatory period\label{operiod} 
		of the set $A$}, and is denoted by $\mathbf{p}$.
It is closely related to the definition of the oscillatory period of lattice self-similar sets studied in \cite{lapidusfrank12}.

		\medskip
		
     $\bullet$ $A$ is a {\em nonperiodic set}\label{nonperiodic} if any function $F(t)$ appearing in (\ref{F}) cannot be written 
     in the form $F(t)=G(\rho(t))$ for all positive $t$ small enough, where $G$ is periodic
     and $\rho$ satisfies conditions (\ref{rho}).

			
\bigskip

\noindent 
Nonperiodic sets can be further classified as follows.
Let $A$ be a nonperiodic set of~$\eR^N$.

\medskip

     $\bullet$\label{kvaziperiodset} $A$ is a {\em transcendentally} (resp., {\em algebraically}) {\em $n$-quasiperiodic set},\label{quasi_periodic_set}
		where $n\ge2$ or $n=\ty$,
		if $F(t)=G(\rho(t))$, where the function $G=G(\tau)$ is transcendentally (resp., algebraically) $n$-quasiperiodic (in the sense of Definition~\ref{quasiperiodic} for $n\ge2$
		and in the sense of Definition \ref{quasipty} for $n=\ty$)
     and the auxiliary function $\rho$ satisfies conditions~(\ref{rho}). Several examples of such sets have been studied in Section~\ref{qp_sets} above as well as in \cite{fzf} and~ [LapRa\v Zu2--5].
		
		\medskip
		
     $\bullet$ $A$ is a {\em nonquasiperiodic set}\label{nonqp}  if it is not quasiperiodic, that is, if any function $F(t)$ appearing in
     (\ref{F}) cannot be written in the form $F(t)=G(\rho(t))$, with $G=G(\tau)$ being quasiperiodic (see Definition~\ref{quasip}) and $\rho$ satisfying conditions~(\ref{rho}).

\bigskip

\section{Perspectives and selected open problems}\label{open}

In the present section, we propose to investigate several open problems and directions for further research, either in connection with functional analysis, dynamical systems and differential (or partial differential) equations (as in the first part of Section~\ref{open}, see Problems~8.1--8.5) or in connection with fractal geometry, complex and conformal dynamics, harmonic analysis and spectral geometry (as in the second part of Section~\ref{open}, see Problems~8.6--8.8).
The interested reader will find a much greater variety of open problems and many other proposed directions of further research throughout~\cite{fzf} and, especially, in the concluding comments chapter of \cite{fzf}, as well as in [LapRa\v Zu2--7]. 
We note, however that the open problems suggested in the first part of the present section (namely, Problems~8.1--8.5) are original to the present survey article.

\medskip

It is natural to study the qualitative properties of solutions of various classes of differential equations, as well as of the trajectories of dynamical systems (both continuous and discrete) and of their corresponding limit sets, from the point of view of the classification of the bounded subsets of Euclidean spaces proposed in the preceding section (Section~\ref{class}). In particular, it is natural to investigate, in this spirit, the new qualitative properties of the {\em graphs} of the solutions of various classes of differential equations.
More precisely, we propose the following open problems along these lines.

\medskip

\begin{problem}\label{ds}
Is there a polynomial vector field possessing a transcendentally $n$-quasiperiodic limit set for a given $n\ge2$ or for $n=\ty$?
A similar question can be asked for discrete dynamical systems as well.
\end{problem}

\medskip

\begin{problem}
Can the limit set of a polynomial vector field in an Euclidean space be (maximally) hyperfractal?
A similar question can be asked for discrete dynamical systems.
\end{problem}

\medskip

\begin{problem}\label{sobolev}
Prove or disprove that there is a real-valued Sobolev function, the graph of which possesses any of the following properties:

\medskip

$(a)$ Minkowski degenerate,

\medskip

$(b)$ quasiperiodic (possibly transcendentally $n$-quasiperiodic with $n\ge2$ or $n=\ty$),

\medskip

$(c)$ (maximally) hyperfractal.
\end{problem}

\medskip

\begin{problem}\label{ss}
Analogous questions can be asked for the {singular sets} of Sobolev functions instead of their graphs. A point $a\in\eR^N$ is said to be a {\em singular point} of a Lebesgue measurable function $f:\eR^N\to\ov\Ce$ if there exist two positive constants $C$ and $\gamma$ such that $|f(x)|\ge C|x-a|^{-\gamma}$ (Lebesgue) almost everywhere in a neighborhood of $a$. The {\em singular set} of $f$ is then defined as the set of all of its singular points.
\end{problem}

\medskip

\begin{problem}\label{pde}
Are there boundary value problems (say, linear elliptic boundary value problems of the second order), which generate weak solutions possessing any of the properties indicated in Problem \ref{sobolev} or Problem \ref{ss}?
\end{problem}

\medskip

Many additional problems, of varying difficulty and dealing with the theory of complex dimensions of fractal sets, can be found in the concluding comments chapter of \cite{fzf}.
We only mention three of these open problems here, in condensed form:

\medskip

\begin{problem}\label{8.5}
Use the fractal zeta functions (that is, the distance and tube zeta functions) introduced in~[LapRa\v Zu1--5] 
(and discussed in the present survey article) to determine the complex dimensions of various classes of fractals, including deterministic and random self-similar fractals, the Devil's staircase (i.e., the graph of the Cantor--Lebesgue curve), the Weierstrass nowhere differentiable curve, the Peano curve, quasidisks, random fractals, as well as the limit sets of Fuchsian and Kleinian groups (naturally occurring in geometric group theory and conformal dynamics), and the classic fractals encountered in complex dynamics, especially Julia sets and the Mandelbrot set.
\end{problem}

\medskip

\begin{problem}\label{8.7}
$(i)$ Determine the {\em spectral} complex dimensions (i.e., the poles of the spectral zeta functions) of various classes of fractal drums, including deterministic and random self-similar fractal drums (see, e.g., [Lap1--3, 
Ger, GerSc] and the relevant references therein), including the Koch snow\-flake drum and its natural generalizations (see, for example, \cite{Lap1,Lap3,LaNeReGr,LaPa,vBGi}), the Devil's staircase drum (defined as the region limited by the staircase and lying above the $x$-axis), quasidisks~\cite{mazja,pommerenke}, connected Julia sets and the Mandelbrot set~\cite{Man}.

\medskip

$(ii)$ Address the exact same problem as in part~$(i)$, except for the {\em geometric} (instead of the spectral) complex dimensions, defined as the (visible) poles of the tube (or distance if the (upper) Minkowski dimension of the given set is strictly less than the dimension of the ambient space) zeta function of relative fractal drums associated with the above fractals.\footnote{In the case of self-similar sets or drums, it may be helpful for the reader to refer to Remark~\ref{3.11/2} above.} 
\end{problem}

\medskip

\begin{problem}\label{8.8}
Obtain fractal tube formulas (in the sense of [Lap--vFr1--3, 
 LapPe1--3, 
LapPeWi1--2]), 
expressed in terms of the corresponding (geometric) complex dimensions (defined as the poles of the associated fractal zeta functions, as in Problem~\ref{8.5}), for the various classes of fractals considered in Problem~\ref{8.5}.\footnote{See, especially, Chapters~5 and~8 as well as Section~13.1 of~\cite{lapidusfrank12}  for the case of fractal strings as well as of certain fractal sprays.
For more general fractal sprays, including self-similar tilings, see~[LapPe2--3] 
and~[LapPeWi1--2] 
(as described in~\cite[Section~13.1]{lapidusfrank12}).} 
\end{problem}

\medskip

\begin{remark}\label{8.81/2}
We refer to Remark~\ref{ftff} above for a positive resolution of Problem~\ref{8.8}, recently obtained by the authors in [LapRa\v Zu6--7]. 
Indeed, as was briefly explained in Remark~\ref{ftff}, in \cite{crasext_ftf} (announced in \cite{cras_ftf}) are obtained, under suitable assumptions, ``fractal tube formulas'' (roughly, expressed in terms of sums over the residues of the tube or the distance zeta functions at each of the visible complex dimensions).
These pointwise and distributional tube formulas (with or without error terms) are valid for arbitrary bounded subsets of $\eR^N$ $(N\geq 1)$ satisfying essentially the same languidity conditions as for the corresponding tube formulas obtained for fractal strings in [Lap--vFr1--3]. 
(See, in particular, \cite[Section~8.1]{lapidusfrank12}.)

\bigskip

In order to fully solve Problem~\ref{8.8}, however, one still has to resolve part $(ii)$ of Problem~\ref{8.7} just above for specific classes of fractals in order to deduce from the general results of [LapRa\v Zu6--7] 
concrete expressions for the fractal tube formulas associated with those classes of fractals.
For example, for (deterministic) self-similar sets (and associated relative fractal drums), one would expect to obtain fractal tube formulas analogous to those obtained for self-similar strings in \cite[Section~8.4]{lapidusfrank12} and later, for a certain class of higher-dimensional self-similar tilings, in [LaPe2--3] 
and, especially, [LapPeWi1--2], 
as described in \cite[Section~13.1]{lapidusfrank12}.\footnote{See also \cite{DeKoOzUr} for an alternative (but similar) proof of a special case of the aforementioned results for self-similar tilings, obtained since then independently of the work in [LapRa\v Zu6--7].
}

\medskip

We should note, nevertheless, that the results of [LapRa\v Zu6--7] 
can be applied to yield concrete fractal tube formulas for a variety of self-similar examples, including the Cantor set (and string), the Sierpi\'nski gasket and its higher-dimensional analogs, the Sierpi\'nski carpet, as well as a three-dimensional analog of the Sierpi\'nski carpet.
Examples of fractals which are not self-similar but for which the concrete (distributional) tube formulas can be derived include ``fractal nests'' and (possibly unbounded) ``geometric chirps''.
(See \cite{fzf} for these notions and [LapRa\v Zu6--7], 
for the corresponding tube formulas.)

\end{remark}





\bigskip

\bigskip

\bigskip

\bigskip

\bigskip

\bigskip

\bigskip

\bigskip

\hbox to\hsize{\vbox to4cm{%
{\obeylines\parindent=0pt\small%
Michel L.\ Lapidus
University of California
Department of Mathematics
900 University Avenue
231 Surge Building
Riverside, CA 92521-0135
USA
{\tt lapidus@math.ucr.edu}
}\vfill}%
\kern-6cm%
\vbox to4cm{\vskip-1mm%
{\obeylines\parindent=0pt\small%
Goran Radunovi\'c and Darko \v Zubrini\'c
University of Zagreb
Faculty of Electrical Engineering and Computing
Department of Applied Mathematics
Unska 3
10000 Zagreb
Croatia
{\tt goran.radunovic@fer.hr} 
{\tt darko.zubrinic@fer.hr}}\vfill
}
}


\end{document}